\documentclass[useAMS,usenatbib]{mn2e}

\usepackage{amssymb}
\usepackage{times}
\usepackage{graphicx}
\usepackage{multirow}

\title[Simulations of the Galactic Polarized Foreground]
{Realistic Simulations of the Galactic Polarized Foreground:
Consequences for 21-cm Reionization Detection Experiments}

\author[Jeli{\'c} et al. ]{Vibor Jeli{\'c}$^{1,2}$\thanks{E-mail:
vjelic@astro.rug.nl}, Saleem Zaroubi$^{1}$, Panagiotis
Labropoulos$^{1}$, Gianni Bernardi$^3$, \newauthor A. G. de Bruyn$^{1,2}$ 
and L\'{e}on V. E. Koopmans$^{1}$ \\ $^{1}$Kapteyn Astronomical
Institute, University of Groningen, P.O. Box 800, 9700 AV Groningen,
the Netherlands \\ $^2$ASTRON, P.O. Box 2, 7990 AA Dwingeloo, the
Netherlands\\ $^{3}$Harvard Smithsonian Center for Astrophysics, 60 
Garden Street, Cambridge, MA 02138, USA}

\begin{document}


\pagerange{\pageref{firstpage}--\pageref{lastpage}} \pubyear{2009}

\maketitle

\label{firstpage}

\begin{abstract}
Experiments designed to measure the redshifted 21~cm line from the Epoch
of Reionization (EoR) are challenged by strong astrophysical
foreground contamination, ionospheric distortions, complex
instrumental response and other different types of noise (e.g. radio
frequency interference). The astrophysical foregrounds are dominated
by diffuse synchrotron emission from our Galaxy. Here we present a simulation
of the Galactic emission used as a foreground module for the
LOFAR- EoR key science project end-to-end simulations. The simulation 
produces total and polarized intensity over $10^\circ \times
10^\circ$ maps of the Galactic synchrotron and free-free emission,
including all observed characteristics of the emission: spatial
fluctuations of amplitude and spectral index of the synchrotron
emission, together with Faraday rotation effects. The importance of
these simulations arise from the fact that the Galactic
polarized emission could behave in a manner similar to the EoR signal
along the frequency direction. As a consequence, an improper
instrumental calibration will give rise to leakages of the polarized
to the total signal and mask the desired EoR signal. In this paper
we address this for the first time through realistic simulations.
\end{abstract}


\section{Introduction}\label{sec:intro}
The Epoch of Reionization (EoR) is expected to occur between redshift 6 and 12, 
as indicated from observed comic microwave background (CMB) polarization
\citep{komatsu09}, high redshift quasar spectra \citep{fan06}, and the thermal 
history of of the intergalactic medium \citep{theuns02b,theuns02a,bolton10}. At
redshifts, the 21~cm line  from neutral hydrogen is shifted into meter wavelengths and
therefore sets the frequency range of EoR experiments
to the long-wavelength part of the radio spectrum ($\sim 100-200~{\rm MHz}$).

There are several planned and ongoing experiments designed to probe
the EoR through redshifted 21~cm emission
line from neutral hydrogen using radio arrays: GMRT\footnote{Giant
Metrewave Telescope, http://gmrt.ncra.tifr.res.in}, LOFAR\footnote{Low
Frequency Array, http://www.lofar.org}, MWA\footnote{Murchinson
Widefield Array, http://www.mwatelescope.org/}, 21CMA\footnote{21
Centimeter Array, http://21cma.bao.ac.cn/}, PAPER\footnote{Precision
Array to Probe EoR, http://astro.berkeley.edu/$^\sim$dbacker/eor}, and
SKA\footnote{Square Kilometer Array, http://www.skatelescope.org/}.

The low-frequency radio sky at these wavelengths is dominated by diffuse 
synchrotron emission from the Galaxy and integrated emission from extragalactic
sources (radio galaxies and clusters). Although, this foreground
emission is $4-5$ orders of magnitude stronger than the expected
EoR signal, the ratio between their intensity fluctuations on arcmin
to degree scales  measured by interferometers
is `only' $2-3$ orders of magnitude \citep{shaver99}. In addition to the
foregrounds, the EoR experiments are also challenged by understanding
of the instrumental response and ionospheric disturbances to high
precision (Labropoulos , in preparation).

Currently there are numerous efforts to simulate all the data
components of the EoR experiments:  cosmological 21~cm signal,
foregrounds, ionosphere and instrumental response. The main aim of 
these end to end simulations is to develop a robust signal extraction 
scheme for the extremely challenging EoR observations \citep[e.g.][]
{santos05, morales06, wang06, jelic08,
bowman09, harker09a, harker09b, panos09}.
 
The foregrounds in the context of the EoR measurements have been
studied theoretically by various authors. \citet{shaver99} have given
the first overview of the foreground components. \citet{dimatteo02, dimatteo04} 
have studied emission from unresolved
extragalactic sources at low radio frequencies. \citet{oh03} and
\citet{cooray04} have considered the effect of free-free emission from
extragalactic haloes. \citet{santos05} carried out a detailed study of 
the functional form of the foreground correlations. \citet{jelic08} have made 
the first detailed foreground 
model and have simulated the maps that include both the diffuse emission 
from our Galaxy and extragalactic 
sources (radio galaxies and clusters). \citet{gleser08} have also studied both
galactic and extragalactic foregrounds. \citet{deoliveira08} has used all
publicly available total power radio surveys to obtain 
all-sky Galactic maps at the desired frequency range and \citet{bowman09} has
studied foreground contamination in the context of the power spectrum estimation. 

Recently, a Galactic 3D emission model has been developed by
\citet{sun08, waelkens09, sun09}
(the \textsc{hammurabi}\footnote{http://www.mpa-garching.mpg.de/hammurabi/}
code), derived from a 3D distribution of the Galactic thermal electrons,
cosmic-ray electrons and magnetic fields. The code is able to reproduce
all-sky or zoom-in maps of the Galactic emission over a wide frequency
range.
 
In addition to simulations, a number of observational projects have 
given estimates of Galactic foregrounds in small selected areas.
\citet{ali08} have used 153 MHz observations with GMRT
to characterize the visibility correlation function of the
foregrounds. \citet{rogers08} have measured the spectral index of the
diffuse radio background between 100 and 200~{\rm MHz}. \citet{pen09}
have set an upper limit to the diffuse polarized Galactic emission; and
\citet{bernardi09a,bernardi10} 
obtained the most recent and comprehensive targeted observations 
with the Westerbork Synthesis Radio Telescope (WSRT).

However, current observations are not able to fully constrain 
the foregrounds, especial the Galactic polarized synchrotron emission, 
as required by EoR experiments. The importance of the polarized foreground
stems from the fact that the LOFAR instrument, in common with all current 
interferometric EoR experiments, has an instrumentally polarized response.
An improper polarization calibration will give rise to leakages of the complex 
polarized signal to the total signal.  Since the Galactic polarized emission is quite
structured along the frequency direction, the leakage of polarized intensity will
have similar structures along the frequency and will mimic the EoR signal. 
Therefore, for reliable detection of the EoR signal it is essential at this stage:
(i) to simulate the polarized foregrounds, and (ii) to test the influence of the leakages 
on the extraction of the EoR signal. This paper focuses on both aspects.

In our previous foreground model \citep{jelic08}, the total 
intensity Galactic emission maps were obtained from 
three Gaussian random fields. The first two
were for the amplitudes of synchrotron and free-free emission and
the third was for the spectral index of the
synchrotron emission.  The polarized maps were simulated in a similar
way but with added multiple 2D Faraday screens along the line of
sight. Despite the ability of that model to simulate observed
characteristics of the Galactic emission (e.g. spatial and frequency
variations of brightness temperature and its spectral index), the
model had some disadvantages: e.g. the Galactic emission was derived
\textit{ad hoc} and depolarization effects were not taken into account.

Our new foreground model, presented in this paper, is  
an extension of our previous foreground model \citep{jelic08}. It simulates
both Galactic synchrotron and free-free emission in total and polarized 
intensity, but in a more realistic way and with some additional benefits.
The Galactic emission in our current model is derived from the
physical quantities and 3D characteristics of the Galaxy
\citep[the cosmic ray and thermal electron density, and the magnetic field; e.g.]
[and references therein]{beck96, berkhuijsen06, sun08}.  In addition,
the model has the flexibility to simulate any peculiar case of the 
Galactic emission including very complex 3D polarized structures 
produced by ``Faraday screens'' and depolarization due to Faraday 
thick layers.

Our Galactic emission model has some similarities with the 
\textsc{hammurabi} model, but the difference between the two 
is the main purpose of the simulations.
The \textsc{hammurabi} simulation is based on a very complex 
Galactic model with aim to reproduce the observed all-sky
maps of the Galactic emission. Because of its complexity, the high
resolution zoom-in maps require a lot of computing power and time
\citep{sun09}. In contrast, our model is restricted to produce fast and
relatively small maps of Galactic emission, which are then used as
a foreground template for the LOFAR-EoR end to end simulation. Since 
the foreground subtraction is usually done along the frequency 
direction, our model also includes 3D spatial variations of the 
spectral index of the Galactic synchrotron radiation.

The paper is organized as follows. Section~\ref{sec:theory} gives a
brief theoretical overview of the Galactic emission and Faraday
rotation. The observational constrains of the Galactic emission are
presented in Sec.~\ref{sec:obs}. The simulation algorithm is described
in Sec.~\ref{sec:sim}, while a few examples of simulated maps 
of the Galactic emission are presented in Sec.~\ref{sec:ex}. Section~\ref{sec:ex}
also give a quantitative comparison of our simulations with the observations.
Section~\ref{sec:lofar-eor} describes the EoR signal and instrumental
response simulations of the LOFAR-EoR pipeline. We
discuss the influence of the polarized foregrounds on the extraction 
of the EoR in Sec.~\ref{sec:calibration}. The paper concludes with summary and
conclusions (Sec.~\ref{sec:so}).

\section{Theory}\label{sec:theory}
In radio astronomy, at frequencies where the Rayleigh-Jeans law is applicable, the
radiation intensity, $I$ (energy emitted per unit time per solid angle
and per unit area and unit frequency), at the frequency $\nu$ is commonly 
expressed in terms of the brightness temperature ($T_b$):
\begin{equation}
T_b(\nu)=\frac{c^2}{2 k_B \nu^2}I(\nu),
\end{equation}
where $c$ is the speed of light and $k_B$ Boltzmann's constant. 

The emission coefficient, $j$ (energy emitted per unit time per solid
angle and per unit volume), at a certain frequency can also be
expressed in terms of the unit temperature, $j_{b}(\nu)=\frac{c^2}{2
k_B \nu^2}j(\nu)$, so that:
\begin{equation}\label{eq:integTb}
T_{b}(\nu)=\int j_{b}(\nu){\rm d}s,
\end{equation}
where the integral is taken along the line of sight (LOS).

In the following subsection we will give a brief theoretical overview of
the Galactic synchrotron and free-free emission, as well as 
Faraday rotation, that will be used later in the simulation. 
The Galactic emission will be
expressed in terms of $j_b$ and $T_b$.
 
\subsection{Synchrotron emission}
Synchrotron emission originates from the interaction between
relativistically moving charges and magnetic fields. In our own galaxy,
synchrotron emission arises from cosmic ray (CR) electrons produced
mostly by supernova explosions and the Galactic magnetic field. A
fairly complete exposition of the synchrotron emission theory is
presented in e.g. \citet{pacholczyk70} and \citet{rybicki86}. Here
we only give a simple description of the emission.

The Galactic synchrotron emission is partially linearly
polarized. Its properties depend on the spatial and energy distribution
of the CR electrons, and the strength and orientation of the
perpendicular (with respect to the LOS) component of the Galactic
magnetic field, $B_{\perp}$.  The emission coefficients of the
Galactic total and polarized synchrotron radiation, $j_b^{Isyn}$ and
$j_b^{PIsyn}$, are given respectively in $cgs$ units, at the frequency $\nu$,
by:
\begin{equation}\label{eq:syn}
j_b^{I,PIsyn}=C_{I,PIsyn}\left( \frac{2\pi m_e
c}{3e}\right)^{-\frac{p-1}{2}}n_{\textsc{cr}}
B_{\perp}^{\frac{p+1}{2}}\nu^{-\frac{p+3}{2}},
\end{equation}
with
\begin{equation}
C_{Isyn}=\frac{\sqrt{3}e^3}{8\pi m_e k_B (p+1)} \Gamma \left(
\frac{p}{4}-\frac{1}{12} \right) \Gamma\left(
\frac{p}{4}+\frac{19}{12} \right),
\end{equation}
\begin{equation}
C_{PIsyn}=\frac{\sqrt{3}e^3}{32\pi m_e k_B} \Gamma \left(
\frac{p}{4}-\frac{1}{12} \right) \Gamma\left( \frac{p}{4}+\frac{7}{12}
\right).
\end{equation}
The charge of the electron is given by $e=4.8\cdot10^{-10}~{\rm Fr}$,
the mass by $m_e=9.1\cdot10^{-28}~{\rm g}$ and $n_{\textsc{cr}}$ is the CR
electron density. Note that for the CR electrons we assume that
their energy spectrum is a power law with a spectral index $p$:
$N(\gamma){\rm d}\gamma=n_{\textsc{cr}0}\gamma^{-p}$, where $\gamma$
is the Lorenz factor and $N(\gamma)$ the number density of electrons
with energy between $\gamma$ and $\gamma+{\rm d}\gamma$ and
$n_{\textsc{cr}0}$ normalization constant. Furthermore, we assume that
their velocity and pitch angle distribution is isotropic.
Both simplifications are consistent with observations and are widely used
by many authors \citep[e.g.][ as most recent examples]{sun08,
waelkens09}. Note that the intrinsic degree of polarization of
synchrotron radiation depends on the energy spectral index $p$ and is
given by
\begin{equation}\label{eq:intpol}
\Pi=\frac{p+1}{p+7/3}.
\end{equation}

The Stokes $Q$ and $U$ parameters of the polarized Galactic synchrotron emission are given
by:
\begin{equation}\label{eq:jQ}
j_b^{Q}=j_b^{PIsyn}\cos{2\Phi},
\end{equation}
\begin{equation}\label{eq:jU}
j_b^{U}=j_b^{PIsyn}\sin{2\Phi},
\end{equation}
where $\Phi$ is polarization angle defined with respect to the 
orientation of the magnetic field.

By integrating $j_b^{Isyn}$, $j_b^{Q}$ \& $j_b^{U}$ along some LOS
(see Eq.~\ref{eq:integTb}) we get the total and polarized Galactic
synchrotron emission in terms of the brightness temperature
($T_b^{Isyn}$, $T_b^{Q}$ \& $T_b^{U}$). Note that observed polarized 
emission and polarization angle $\Phi_{obs}$ are given by:
\begin{equation}\label{eq:obsI}
T_b^{PI}=\sqrt{(T_b^Q)^2+(T_b^U)^2},
\end{equation}
\begin{equation}\label{eq:obsPHI}
\Phi_{obs}=\frac{1}{2}\arctan\frac{T_b^U}{T_b^Q}.
\end{equation}

\subsection{Free-free emission}
Radiation due to (de)acceleration of a charged particle in the
electrical field of another is called bremsstrahlung or free-free
radiation. The Galactic free-free emission originates from
electron-ion encounter in the warm ionized gas. As for the synchrotron
emission a detail theory of the free-free radiation can be found in
e.g. \citet{rybicki86} and \citet{wilson09}, here we give only the
necessary formulae.

The optical depth, $\tau_{\nu}^{ff}$, of the warm ionized gas at a
given low radio frequency $\nu$ is:
\begin{equation}\label{eq:tauff}
\tau_{\nu}^{ff}=3.01\cdot10^{4}g_{ff}\left(\frac{T_e}{[\rm
K]}\right)^{-\frac{3}{2}}\left(\frac{\nu}{[\rm
MHz]}\right)^{-2}\frac{EM}{[\rm cm^{-6}pc]} ,
\end{equation}  
where $T_e$ is temperature of the ionized gas, $g_{ff}$ is the Gaunt
factor of the free-free transition given by:
\begin{equation}
g_{ff}=\ln\left[49.5\left(\frac{\nu}{[\rm
MHz]}\right)^{-1}\right]+1.5\ln\left(\frac{T_e}{[\rm K]}\right),
\end{equation}
and $EM$ is emission measure defined as:
\begin{equation}
\frac{EM}{[\rm cm^{-6}pc]}=\int n_e^2{\rm d}s.
\end{equation}
The integral is taken over the LOS, where $n_e$ in $\rm cm^{-3}$ is
the electron density of the warm ionized gas.

The Galactic free-free emission in terms of brightness temperature,
$j_b^{ff}$, is given by:
\begin{equation}\label{eq:ff}
j_b^{ff}=T_e(1-{\rm e}^{-\tau^{ff}}).
\end{equation}
Note that for optically thin ionized gas, $j_b^{ff}$ is simply
given by $j_b^{ff}=T_e\tau^{ff}$.
\subsection{Faraday rotation}
When the polarization angle of an electromagnetic wave is rotated
while passing through a magnetized plasma, the effect is called
Faraday rotation \citep[for details see][]{rybicki86, wilson09}. The
rotation depends on the frequency of the wave, $\nu$, electron
density, $n_e$, and magnetic field component parallel to the LOS,
 $B_{\parallel}$:
\begin{equation}\label{eq:FR}
\Phi=\Phi_0+\frac{e^3}{2\pi m_e^2c^2}\nu^{-2}\int n_e B_{\parallel}
{\rm d}s,
\end{equation}
where the polarization angle of the wave before rotation is denoted with
$\Phi_0$.  Eq.~\ref{eq:FR} is also written as 
$\Phi=\Phi_0+RM\lambda^2$ with $\lambda$ in units of ${\rm
m}$ and RM (rotation measure) defined as:
\begin{equation}\label{eq:RM}
\frac{RM}{[\rm rad~m^{-2}]}=0.81\int\frac{n_e}{[\rm
cm^{-3}]}\frac{B_{\parallel}}{\rm [\mu G]}\frac{{\rm d}s}{\rm [pc]}.
\end{equation}
The RM is positive when $B_{\parallel}$ points towards observer
and negative when $B_{\parallel}$ points in away.

\section{Observational constraints}\label{sec:obs}
There are several all-sky maps of the total Galactic diffuse radio
emission at different frequencies and angular resolutions
\citep{haslam82, reich86, reich88, page07}. The $150~{\rm MHz}$ map by
\citet{landecker70} is the only all-sky map in the frequency range
($100-200~{\rm MHz}$) relevant for the EoR experiments, but has only
$5^\circ$ resolution.

At high Galactic latitudes the minimum brightness temperature of the
Galactic diffuse emission is about $20~{\rm K}$ at $325~{\rm MHz}$
with variations of the order of 2 per cent on scales from 5 to 30
${\rm arcmin}$ across the sky \citep{debruyn98}. At the same Galactic
latitudes, the temperature spectral index of the Galactic emission is
about $-2.55$ at between 100 and 200 MHz \citep{rogers08}
and steepens towards higher frequencies
\citep[e.g.][]{platania98, bennett03, bernardi04}. 
Furthermore, the spectral index
gradually changes with position on the sky. This change appears to be
caused by a variation in the spectral index along the line of
sight. An appropriate standard deviation in the power law index, in
the frequency range 100--200~{\rm MHz} appears to be of the order of
$\sim 0.1$ (Shaver et al. 1999).

Using the obtained values at $325~{\rm MHz}$ and assuming the
frequency power law dependence, the Galactic diffuse emission is
expected to be $140~{\rm K}$ at $150~{\rm MHz}$, with $\sim 3~{\rm K}$
fluctuations.

Studies of the Galactic polarized diffuse emission are done mostly at
high radio ($\sim 1 {\rm GHz}$) frequencies \citep[for a recent review 
see,][]{reich06}. At lower frequencies ($\sim 350~{\rm MHz}$), there are several
fields done with the Westerbork telescope (WSRT) \citep{wieringa93,
haverkorn03, schnitzler08}. These studies revealed a
large number of unusually shaped polarized small-scale structures of
the Galactic emission, which have no counterpart in the total
intensity. These structures are usually attributed to the Faraday
rotation effects along the line of sight. 

At high Galactic latitudes, the Galactic polarized emission at
$350~{\rm MHz}$ is around $5~{\rm K}$ or more, on 5--10~{\rm arcmin}
scales \citep{debruyn06}. At $150~{\rm MHz}$ this polarized emission
would scale to few tens of Kelvin if it were Faraday thin. However, depolarization,
that is prominent at low radio frequencies, can significantly lower
the level of polarized emission.

Recently, a comprehensive program was initiated by the LOFAR-EoR
collaboration to directly measure the properties of the Galactic radio
emission in the frequency range relevant for the EoR experiments. The
observations were carried out using the Low Frequency Front Ends
(LFFE) on the WSRT radio telescope. Three different fields were
observed. The first field was a highly polarized region known as the ``Fan
region'' in the 2nd Galactic quadrant  at a low Galactic latitude of $\sim10^\circ$
\citet{bernardi09a}. The second field 
was a very cold region in the Galactic halo ($l\sim170^\circ$) around the bright radio quasar
 3C196, and third was a region around the North Celestial Pole \citep[NCP, $l\sim125^\circ$][]  
{bernardi10}. The last two
fields represent possible targets for the LOFAR-EoR observations. Below
we present the main results of these papers.

In the ``Fan region'', fluctuations of the Galactic diffuse emission were 
detected at $150~{\rm MHz}$ for the first time. The fluctuations were detected 
both in total and polarized intensity, with an $rms$ of $14~{\rm K}$ 
($13~{\rm arcmin}$ resolution) and $7.2~{\rm K}$ ($4~{\rm arcmin}$ resolution)
respectively \citep{bernardi09a}. Their spatial structure appeared to
have a power law behavior with a slope of $-2.2\pm0.3$ in total intensity
and $-1.65\pm0.15$ in polarized intensity. Note that, due to its strong 
polarized emission, the ``Fan region'' is not a representative part of the
 high Galactic latitude sky.

Fluctuations of the total intensity Galactic diffuse emission in the
``3C196'' and ``NGP'' fields were also observed on scales larger than
$30~{\rm arcmin}$, with an $rms$ of $3.3~{\rm K}$ and $5.5~{\rm K}$
respectively. 

Patchy polarized emission was found in the ``3C196''
field, with an $rms$ value of $0.68~{\rm K}$ on scales larger than
 $30~{\rm arcmin}$ \citep{bernardi10}.  Thus, 
the Galactic polarized emission fluctuations seem to be smaller than
expected by extrapolating from higher frequency
observations. Recent observations at mid-galactic latitude 
with the Giant Metrewave Radio Telescope (GMRT) confirm this conclusion, 
by setting an upper limit to the diffuse polarized Galactic emission in their field to 
be $< 3~{\rm K}$ at $150~{\rm
MHz}$ and on scales between 36 and 10~{\rm arcmin}  \citep{pen09}.

\section{Simulation}\label{sec:sim}
In this section, the various components of the simulation that lead
towards the brightness temperature maps of the Galactic synchrotron
and free-free emission in a total and polarized intensity are
explained. Because the simulated maps will be used as a foreground
template for the LOFAR-EoR end-to-end simulations, the foreground 
simulations assume the
angular and frequency range of the LOFAR-EoR experiment, i.e.
$10^\circ\times10^\circ$ maps from $115~{\rm MHz}$ to $180~{\rm
MHz}$. In addition, all parameters of the simulations can be tuned to
any desired value or have any desired characteristic, allowing to
explore the parameter space of our Galactic model. 

Our algorithm is based on a 3D grid in a Cartesian coordinate
system, where $xy$-plane represents the angular plane of the sky (``flat
sky'' approximation valid for a small field of view) and $z$ axis is a
line of sight direction in parsecs.

The first step in our simulation is to calculate, at a certain
frequency, the 3D emission coefficient of the Galactic synchrotron and
free-free emission expressed in terms of unit temperature (see
Eq.~\ref{eq:syn} \& \ref{eq:ff}). The emission coefficients are
obtained from the cosmic-ray, $n_{\textsc{cr}}$, and thermal
electron, $n_e$, densities, and the Galactic magnetic field
($\vec{B}$). Given the 3D emission coefficients, we integrate along
the LOS to obtain the brightness temperature maps of the Galactic
synchrotron and free-free emission at a certain frequency. The
calculation also includes Faraday rotation effects. Note that all
parameters of the simulation are set in such a way that simulated 
maps are  quantitatively (e.g. presence of the structures at different 
scales, spatial and frequency variations of the brightness temperature and  
its spectral index, etc.) in agreement with
the observations overviewed in Sec.~\ref{sec:obs}.
A detail description of all input parameters ($n_{\textsc{cr}}$, 
$n_e$ and $\vec{B}$) and the algorithm follow. 
A few examples of simulated Galactic emission data cubes 
are presented in Sec.~\ref{sec:ex}, together with a quantitative 
comparison with the observations.
 
\subsection{Cosmic ray electron density}\label{sec:CRe} 
The cosmic ray (CR) electrons relevant for the Galactic synchrotron
emission have energies between $400~{\rm MeV}$ and $25~{\rm GeV}$,
assuming a Galactic magnetic field of a few $\mu G$
\citep{webber80}. In this energy range, the CR electron distribution
can be described as a power law. The power law is normalized according
to the measurements obtained in the solar neighborhood. However, the
locally measured values might not be a good representative for the CR
density elsewhere in the Galaxy \citep[e.g.][]{strong04}. As a
consequence, the CR electron distribution is weakly
constrained.

In our simulation, uniform CR electron density
distribution is assumed in the $xy$-plane. In the $z$ 
direction we follow \citet{sun08} and assume an exponential 
distribution:
\begin{equation}\label{eq:CR}
  n_{\textsc{cr}}=n_{\textsc{cr}0}\exp{\left( \frac{-z}{1 {\rm kpc}}\right)}.
\end{equation}
Note that $n_{\textsc{cr}0}$ depends on the assumed energy spectral 
index $p$ of the CR electrons, so it is normalized according to
Eq.~\ref{eq:integTb} for the synchrotron radiation. Assuming 
$T_b(150~{\rm MHz})\simeq145~{\rm K}$, $B_\perp=5~{\rm \mu G}$
and $p=2$, we get $n_{\textsc{cr}0}\simeq1.4\cdot10^{-8}~{\rm cm^{-3}}$. 

In the desired frequency range of our simulation, the assumed energy 
spectral index $p=2$ is consistent with the values of the
typically observed brightness temperature spectral index of the
Galactic synchrotron emission ($\beta=-2.5$, see Sec.~\ref{sec:obs}
)\footnote{The brightness
temperature spectral index $\beta$ of the Galactic synchrotron
emission and the energy spectral index $p$ of the CR electrons are
related as $\beta=-(p+3)/2$.}.

In addition, the spatial variations of the spectral index $p$ are
introduced to mimic the observed spatial fluctuations of $\beta$. We
follow our previous model \citep{jelic08} and simulate the variation of
$p$ (or $\beta$) as a Gaussian random field. For the power
spectrum of the Gaussian random field it is assumed a power law with index $-2.7$ \citep{jelic08}.

\subsection{Galactic magnetic field}\label{sec:MF}
The Galactic magnetic field has two components: a regular component
$\vec{B_{r}}$ and a random component $\vec{b}$, so that
the total Galactic magnetic field is given as $\vec{B}=\vec{B_r}+\vec{b}$ 
 \citep[for review see][]{beck96, han02}.
The regular component is usually simulated as a combination of a disk 
and a halo field, whereas the random field component is simulated 
as a Gaussian random field, Gaussian random field, \citep[for details see][]{sun08, sun09}.
Note that for our calculations, we split $\vec{B}$ in a component 
parallel ($B_{\parallel}$) and perpendicular ($B_{\perp}$) to the LOS,
so that   Faraday rotation is defined by $B_{\parallel}$ and 
synchrotron emission by $B_{\perp}$.

Considering the aim of our effort to simulate the Galactic emission
for a small patch of the sky, we treat the regular field component in
a simplified way. The regular field component is assumed to be uniform
in the $xy$-plane and to have an exponential decrease in the $z$
direction. The typical value of the regular field component is a few $\mu
G$ \citep[for review see][]{beck96, han02}.

For the random field component we follow \citet{sun08, sun09} and
simulate it as a Gaussian random field. The power spectrum of
the field follows a power law, with spectral index $-8/3$. This
spectral index is commonly used for a Kolmogorov-like turbulence
spectrum.

In our simulation, the realization of the random field component is
done in the following way. First we generate three different Gaussian random fields for
the $b_x$, $b_y$ and $b_z$ component. From those three
fields, we then calculate the amplitude of $\vec{b}$ and normalize it
to the desired value. A typical value for the mean random field strength
is $b=3~{\rm \mu G}$ \citep{sun08}.

\subsection{Thermal electron density}\label{sec:Te}
At high Galactic latitudes, the warm ionized medium consists mostly
of diffuse ionized gas (DIG) with total emission measure of $\sim
5~{\rm pc~cm^{-6}}$ and $T_e=8000~{\rm K}$ \citep{reynolds90}. The
properties of the DIG can be traced by its free-free emission and
dispersion measure\footnote{The dispersion measure is defined as the
integral of the thermal electron density along the LOS. Knowing the
distance to the pulsars (e.g. determined by parallax), an electron
density model can be obtained by a fit to the observed DMs.} (DM) of
pulsars \citep[e.g.][]{gaensler08}.

Recent simulations of the Galactic emission \citep{sun08,
waelkens09, sun09} used the \citet{cordes02} model for the
thermal electron distribution. That model simulates the Galaxy as
several large-scale (e.g. thin and thick disk, and spiral arms) and
small-scale (e.g. supernovae bubbles) structures. In our simulation,
we follow our previous model of the Galactic free-free emission
\citep{jelic08} and simulate the thermal electron density distribution
as a Gaussian random field with the power law type of the spectrum. The spectral index
of the power law is $-3$. The amplitude of the Gaussian random field (thermal electron
density) is normalized in a way to match the typical observed EM of
the quasars at high Galactic latitudes \citep[EM values are taken
from][]{berkhuijsen06}.

It is important to note that our model is flexible to include
additional features of the thermal electron distribution, e.g. dense
bubbles or clumpy distribution. Some of these features are
presented in Sec.~\ref{sec:ex}.

\subsection{The Algorithm}\label{sec:al}
\begin{figure}
\centering \includegraphics[width=.45\textwidth]{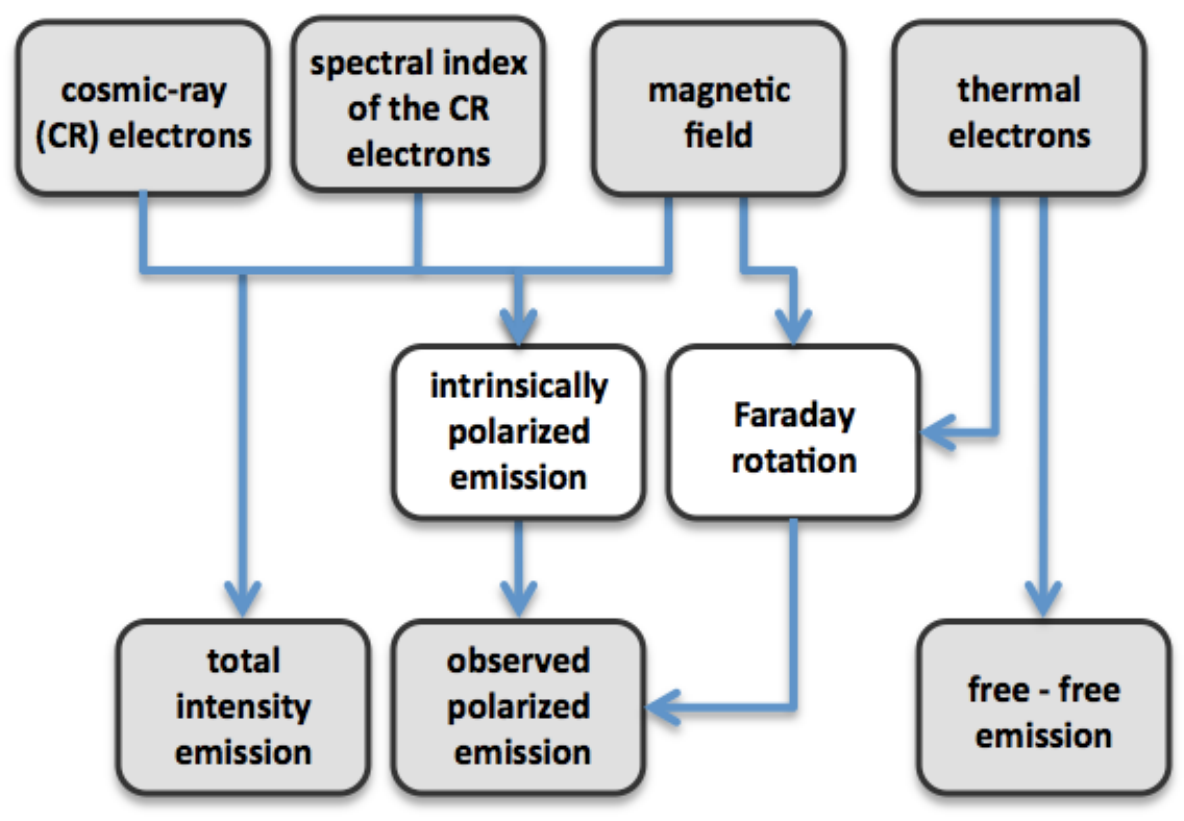}
\caption{Flow chart of the algorithm: the Galactic emission (synchrotron 
and free- free) is derived from the physical quantities and 3D characteristics 
of the Galaxy, i.e. cosmic ray, $n_{\textsc{cr}}$, and thermal electron, $n_e$, 
density; and magnetic field, $\vec{B}$.  In addition,
the algorithm includes Faraday rotation effects.}
\label{fig:fc}
\end{figure}

Here we summarize the steps we follow to obtain maps of the
Galactic emission at a desired frequency. The flow chart of
the algorithm is presented in Fig.~\ref{fig:fc}.

\begin{enumerate}
\item The CR electron density, $n_{\textsc{cr}}$, and the regular
  component, $\vec{B_r}$, of the Galactic magnetic field are defined
  on 3D grid. The distributions of $n_{\textsc{cr}}$ and $\vec{B_r}$
  are uniform in the $xy$-plane and have an exponential decrease
   in the $z$ direction.
\item The spatial distribution of the CR electron energy spectral index, $p$, 
the random component, $\vec{b}$, of the Galactic magnetic
  field and the thermal electron density, $n_e$, are simulated as
  Gaussian random fields. The Gaussian random fields are normalized to result in a desired
  $rms$ value of the brightness temperature maps. Note that additional
  features in the electron distribution are added if desired.
\item The parallel, $B_{\parallel}$, and perpendicular, $B_{\perp}$,
  component of the total Galactic magnetic field, $\vec{B}$, are
  calculated from $\vec{B_r}$ and $\vec{b}$.
\item Using Eq.~\ref{eq:syn}, the emission coefficients of the
  Galactic total, $j_b^{Isyn}$, and polarized,  $j_b^{PIsyn}$, 
  synchrotron radiation are calculated.
\item The optical depth, $\tau^{ff}$, and emission coefficient, 
$j_b^{ff}$, of the thermal plasma are obtained from
  Eq.~\ref{eq:tauff} \& \ref{eq:ff}. Note that these effect is really 
  significant only on the lowest radio frequencies.
\item Absorption of the synchrotron emission by the optical thickness of
 the ionized plasma is taken into account as $\exp{(-\tau^{ff})}$
  factor.
\item Using Eq.~\ref{eq:FR}, the Faraday rotation effect is
  calculated and the polarization angle, $\Phi$, is obtained. Note
  that the intrinsic polarization angle, $\Phi_0$, is defined as the
  inclination of $B_{\perp}$.
\item The Stokes Q and U emission coefficients of the polarized
  emission, $j_b^{Q}$~\&~$j_b^{U}$, are calculated using 
  Eq.~\ref{eq:jQ}~\&~\ref{eq:jU}.
\item By integrating $j_b^{Isyn}$, $j_b^{Q}$~\&~$j_b^{U}$ along some
  LOS (see Eq.~\ref{eq:integTb}), the total and polarized Galactic
  synchrotron emission in terms of the brightness temperature,
   $T_b^{Isyn}$, $T_b^{Q}$ \& $T_b^{U}$, are obtained.
\item Finally, the maps of the total polarized emission ($T_b^{PI}$) and
  observed polarization angle $\Phi_{obs}$ is calculated using
  Eq.~\ref{eq:obsI}~\&~\ref{eq:obsPHI}.
\end{enumerate}
In the following section we will show some examples of the Galactic
emission maps obtained by this algorithm.

\section{Examples of simulated Galactic emission}\label{sec:ex}
Here we demonstrate the ability of our algorithm to realistically
simulate Galactic synchrotron and free-free emission both in total and 
polarized intensity. We also quantitatively compare results of our 
simulations with the observations.
\subsection{Simulated data cubes}
\begin{figure}
\centering \includegraphics[width=.45\textwidth]{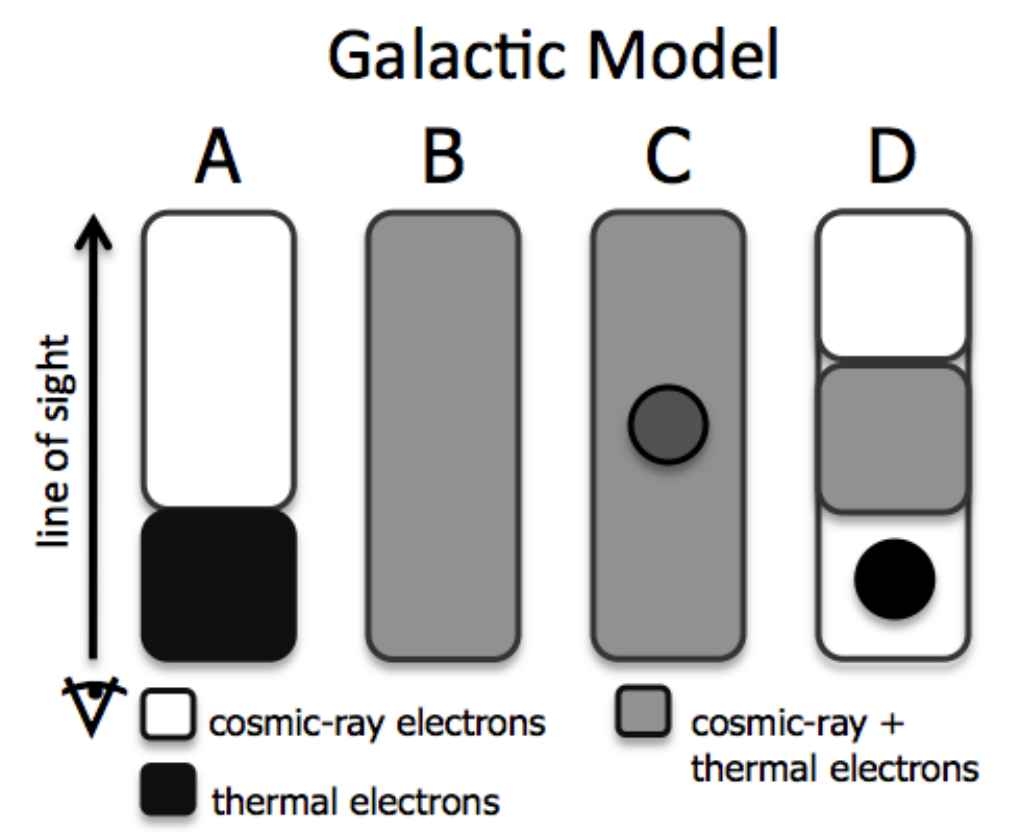}
\caption{The intensity distributions (see
Fig.~\ref{fig:synff}~\&~\ref{fig:PImodels}) are obtained for four
different models of the Galactic emission .  The first (model A) 
assumes that synchrotron and free-free emitters are spatially separated,
 so that thermal plasma acts as a ``Faraday screen''. The second, third 
 and fourth (model B, C, \& D) have regions where both types  permeate 
 in a different way. The synchrotron
emission is differentially Faraday rotated and depolarization occurs.}
\label{fig:models}
\end{figure}

The simulated data cubes are presented 
for four cases of the Galactic emission. First three models show ability 
of our algorithm to simulate different cases of Faraday rotation and 
depolarization, while the last model is tailored to be in agreement with 
\citet{bernardi09a} observations of the Fan region.
\begin{itemize}
\item Model A: CR electrons are distributed in a region of
  $1~{\rm kpc}$ in depth along the LOS ($n_{\rm CR0}=1.4\times10^{-8}~{\rm cm^{-3}}$). 
  In front of this region, there is a thermal electron cloud of $300~{\rm pc}$ in depth along the
  LOS, with an average emission measure of $\bar{EM}=8~{\rm cm^{-6}pc }$, and $T_e=8000~{\rm K}$. The thermal electron cloud is acting as a ``Faraday screen''
  that rotates the polarization angle of synchrotron emission. It is assumed 
  that Galactic magnetic field is uniform throughout the region ($B_{r,\parallel}=3~{\rm \mu G}$ and $B_{r,\perp}=2.5~{\rm \mu G}$).
\item Model B: Both CR and thermal electrons are mixed in
  the region of $1~{\rm kpc}$ in depth along the LOS ($n_{\rm CR0}$, $\bar{EM}$, $T_e$ and $\vec{B_r}$ are the same as in model A). The polarized
  synchrotron radiation is differentially Faraday rotated. Since the 
  Galactic magnetic field has the same characteristics as in model A,
  depolarization is produced only by thermal plasma.
\item Model C: The same as model B, but in the middle of the simulated region there is a dense thermal electron bubble $300~{\rm pc}$ in depth along the LOS, with a strong magnetic field ($B_\parallel=10~{\rm \mu G}$). The Faraday rotation along the bubble will be much larger than in other parts of the region. Note that the size of the bubble is quite large in order to make its appearance in the final maps more clear. Moreover, it is assumed that Galactic magnetic field in not uniform throughout the region: $B_{r,\parallel}$ exponentially decreases along the LOS. We also include a random magnetic field component as a Gaussian random field ($\bar{b_\parallel}=3~{\rm \mu G}$, see Sec.~\ref{sec:sim}). Depolarization is then produced both by thermal plasma and variations of the magnetic field.
\item Model D: This model is based on a proposed cartoon of the magnetized interstellar medium in the Fan region by \citet[][see fig. 12]{bernardi09a}. The intrinsically polarized background emission passes through multiple Faraday screens which make the spatial distribution of the polarized emission more clumpy.  A bubble of ionized thermal plasma sits between the Faraday screens and the observer. This bubble further rotates the polarized emission to the higher values of RM. All parameters of this model are set to match the observed properties of this region: fluctuations of the diffuse emission at $150~{\rm MHz}$ and their power-law type of the power spectrum \citep[see][ and Sec.~\ref{sec:obs}]{bernardi09a}.
\end{itemize}
Cartoons of these four models are given in Fig.~\ref{fig:models}. The Galactic emission maps 
are obtained in the frequency range from $115$ to $180~{\rm MHz}$, with $0.5~{\rm MHz}$
step.

In the first three models the synchrotron emission originates from the same CR electron
distribution with uniform $B_{\perp}$ component of the Galactic magnetic field. Therefore, 
the brightness temperature maps of the total and intrinsic\footnote{Here, the intrinsic polarized
emission, $iPI_{\rm syn}$, means emission defined by Eq.~\ref{eq:syn}. Note that the 
polarization angle of this emission is assumed to be uniform across
the whole region. Thus any effect caused by thermal electrons or $B_{\parallel}$ will
be immediately apparent.} polarized synchrotron emission are
equivalent in all three models. The same is valid for the
free-free emission, i.e., $n_e$ is normalized to the same
value of $EM$. The resulting $10^\circ\times10^\circ$ maps at $150~{\rm
MHz}$ are shown in Fig.~\ref{fig:synff}.a,~b~\&~d, while the $mean$ and
$rms$ of the maps are given in Table~\ref{tab:maps}. A map of the brightness
temperature spectral index $\beta$ of simulated total
intensity synchrotron emission is shown in Fig.~\ref{fig:synff}.c.

Simulated polarized emission maps of the first three Galactic synchrotron
emission models are shown in Fig.~\ref{fig:PImodels}.a, ~b~\&~c. Their $mean$
and $rms$ values together with the degree of polarization and
depolarization are listed in Table~\ref{tab:maps}. A random line through synchrotron 
total and polarized intensity frequency data cubes of first three models are given 
in Fig.~\ref{fig:los}.

Simulated total and polarized emission images of the Fan region (model D) are given
at 150~MHz in Fig.~\ref{fig:Fan} a \& b. For easier comparison with the observation 
an RM image, obtained by applying rotation measure synthesis technique
\citep{brentjens05} on the simulated data, is shown in Fig.~\ref{fig:Fan} c.

\subsection{Discussion and tests of the algorithm}\label{sec:dta}
\begin{figure*}
\centering \includegraphics[width=.24\textwidth]{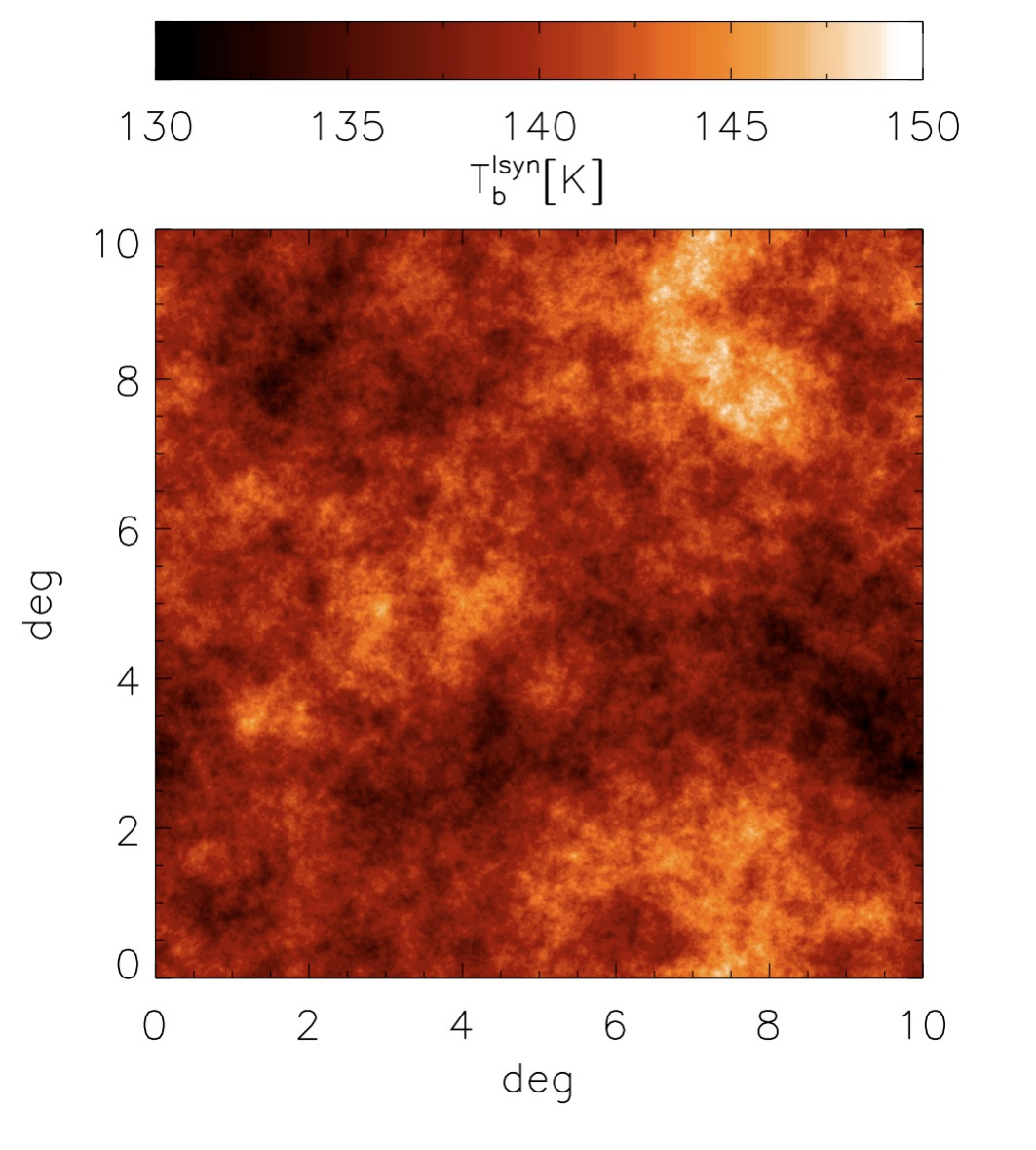}
\centering \includegraphics[width=.24\textwidth]{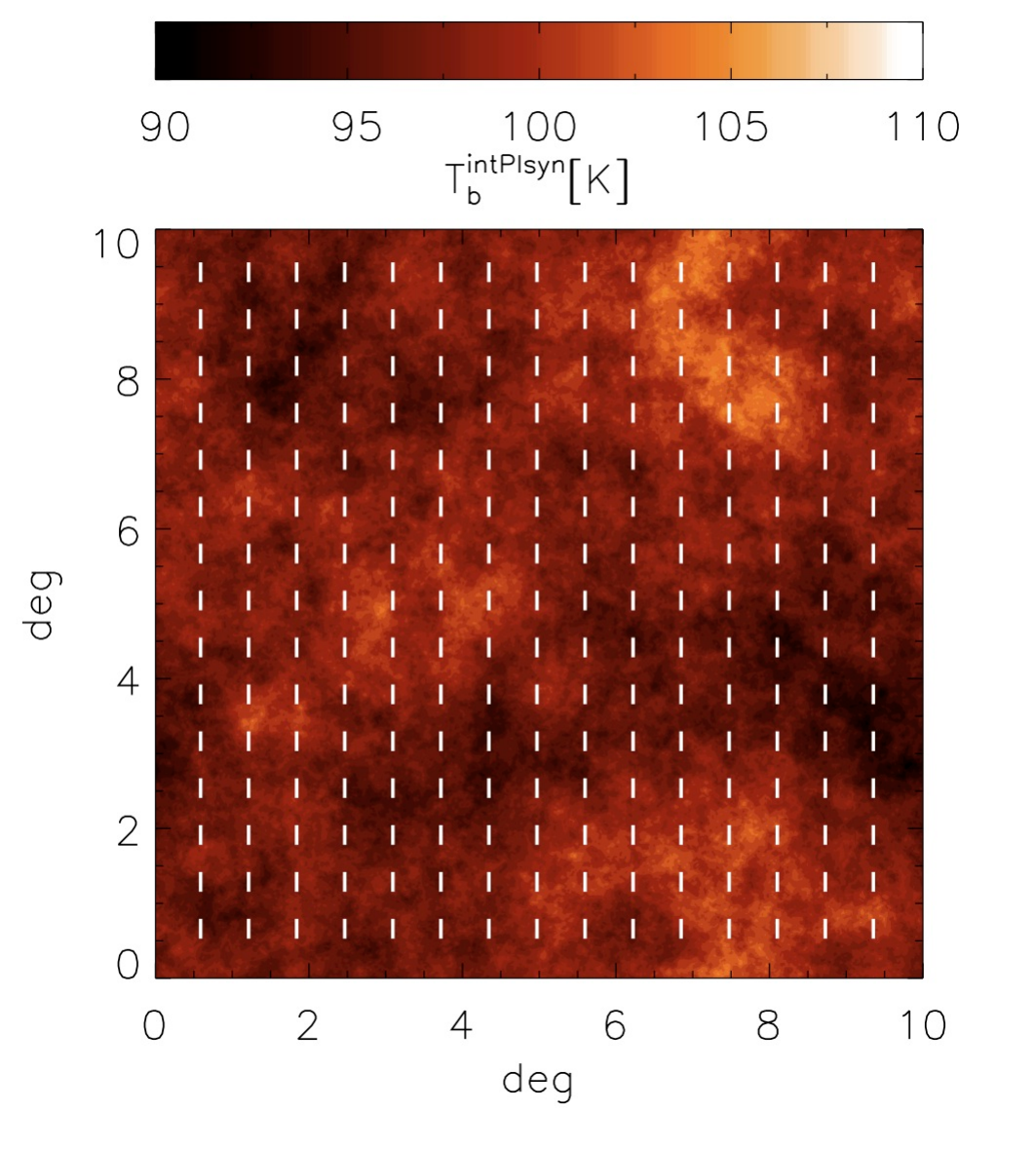}
\centering \includegraphics[width=.24\textwidth]{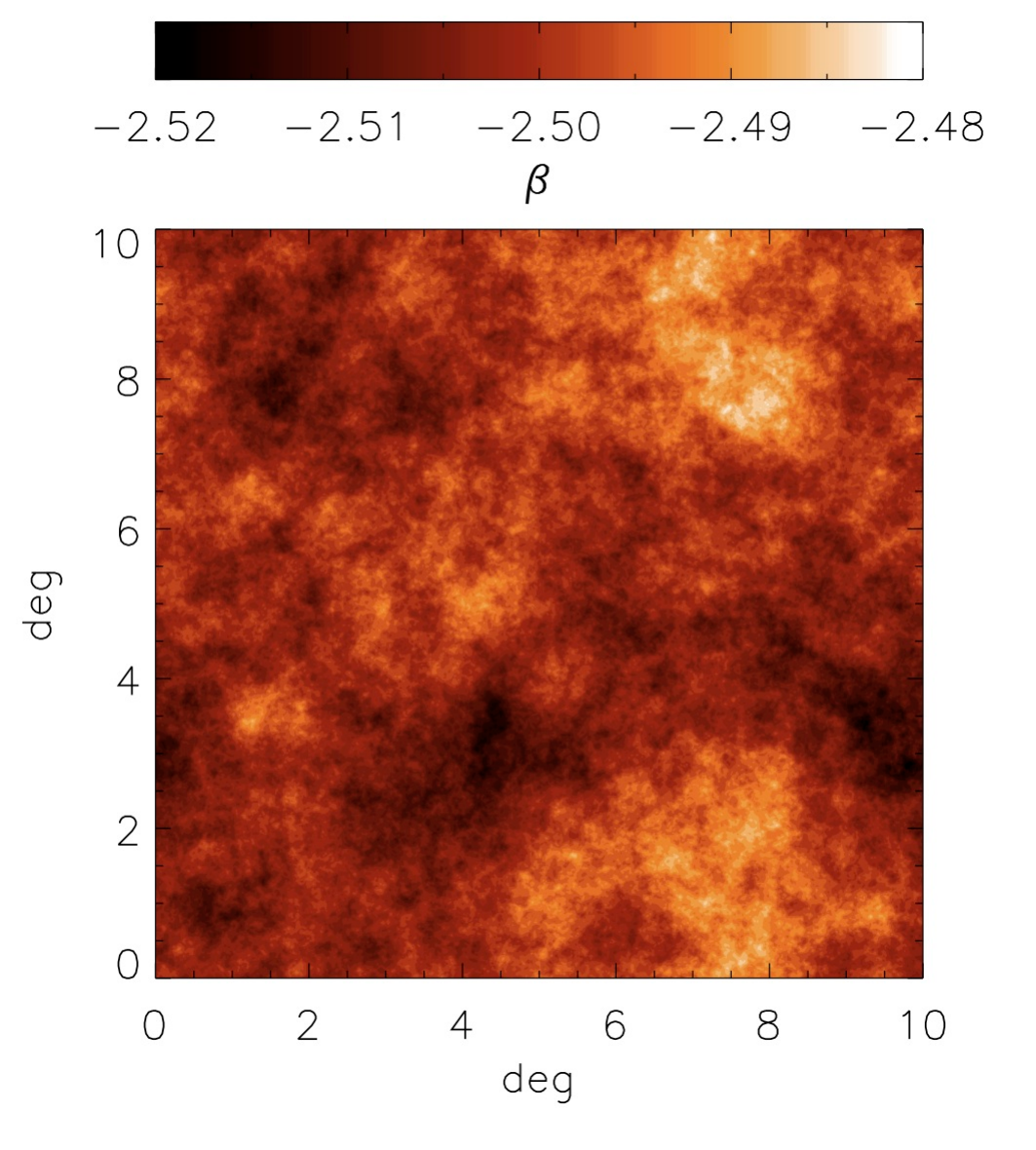}
\centering \includegraphics[width=.24\textwidth]{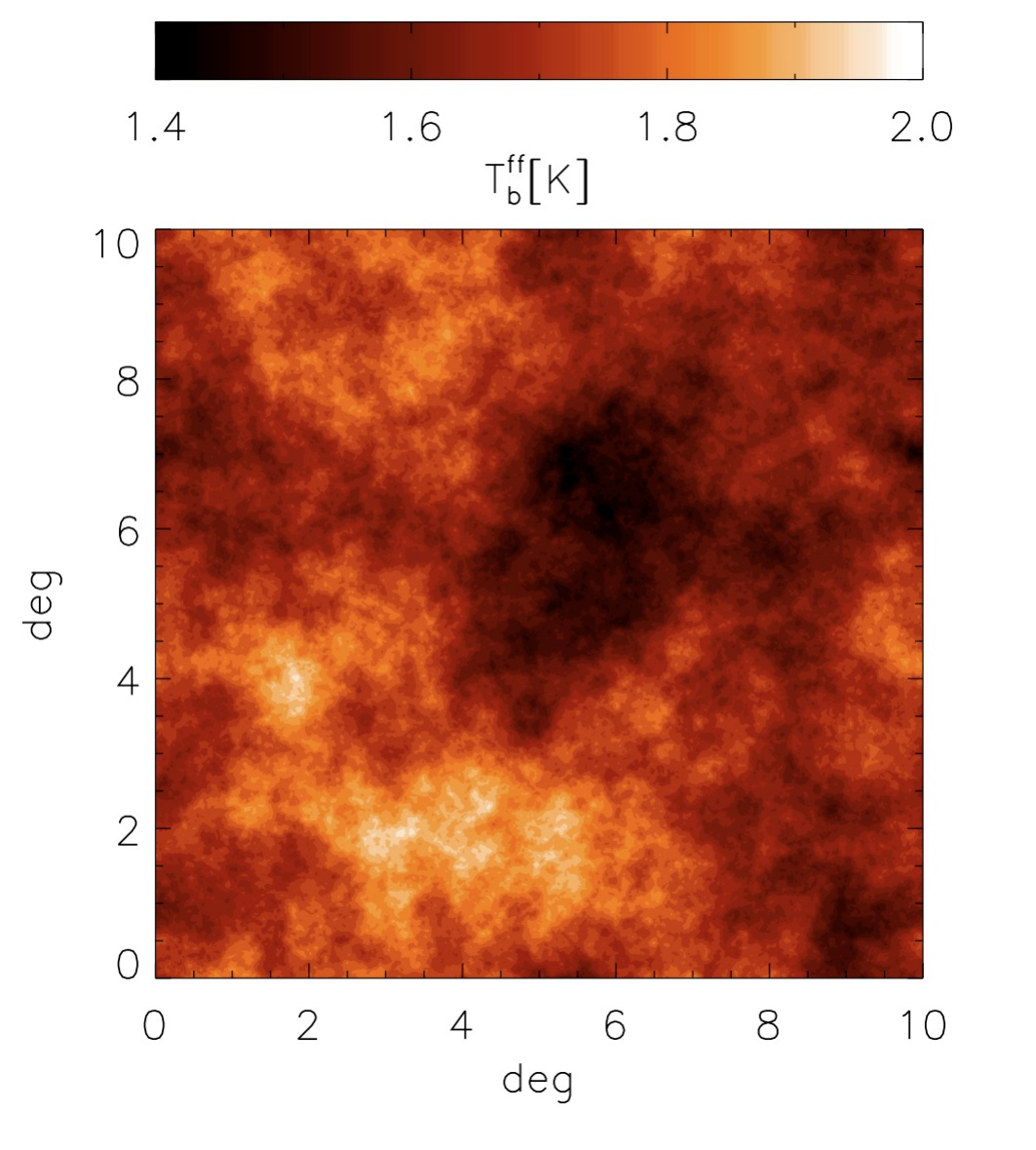}
\\ (a)\hspace{4.cm}(b)\hspace{4.cm}(c)\hspace{4.cm}(d)
\caption{Simulated maps of the total (Fig.~\ref{fig:synff}.a) and
intrinsic polarized (Fig.~\ref{fig:synff}.b) intensity of the Galactic
synchrotron emission. The polarization angle is plotted over the
polarized map as a white lines. The
map of brightness temperature spectral index $\beta$ of simulated total
intensity synchrotron emission is shown in Fig.~\ref{fig:synff}.c.
The total intensity map of the
free-free emission is shown in Fig.~\ref{fig:synff}.d. The angular
size of the maps are $10^\circ\times10^\circ$, with $\sim 1~{\rm
arcmin}$ resolution. The color bar represents the brightness
temperature $T_b$ of emission in kelvin at $150~{\rm MHz}$. The $mean$
and $rms$ value of the maps are given in the Table~\ref{tab:maps}. }
\label{fig:synff}
\end{figure*}

First test of the algorithm is to estimate the degree of
intrinsic polarization, $\Pi$, of the simulated map (Fig.~
~\ref{fig:synff}.b). By 
dividing the intrinsic polarized emission map and the total 
intensity map, we obtain $\Pi=0.69$ (see Table~\ref{tab:maps}). 
This value is in a good agreement with the expected 
theoretical value $\Pi_{p=2}=9/13$ (see Eq.~\ref{eq:intpol}).

\begin{figure*}
\centering \includegraphics[width=.32\textwidth]{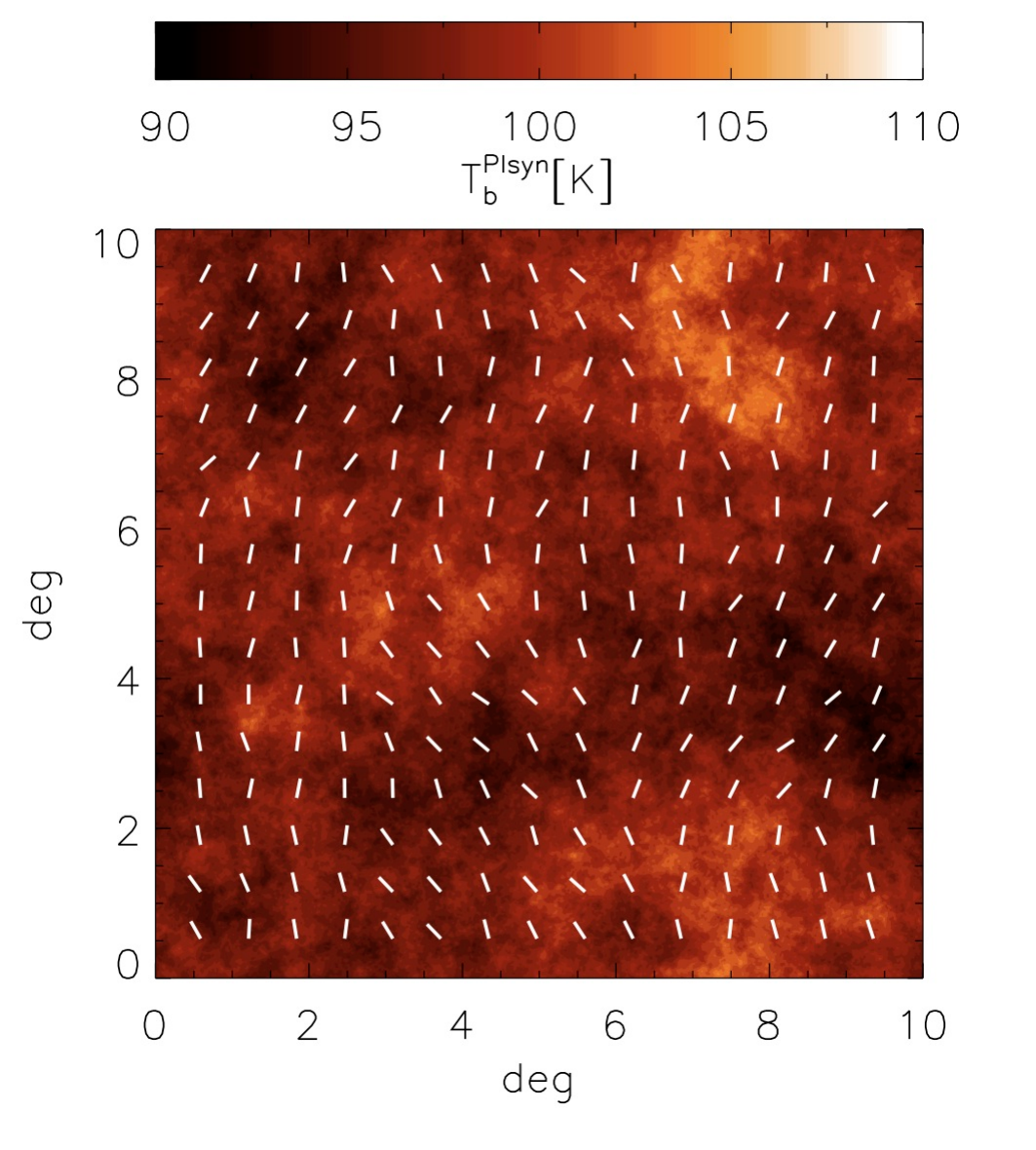}
\centering \includegraphics[width=.32\textwidth]{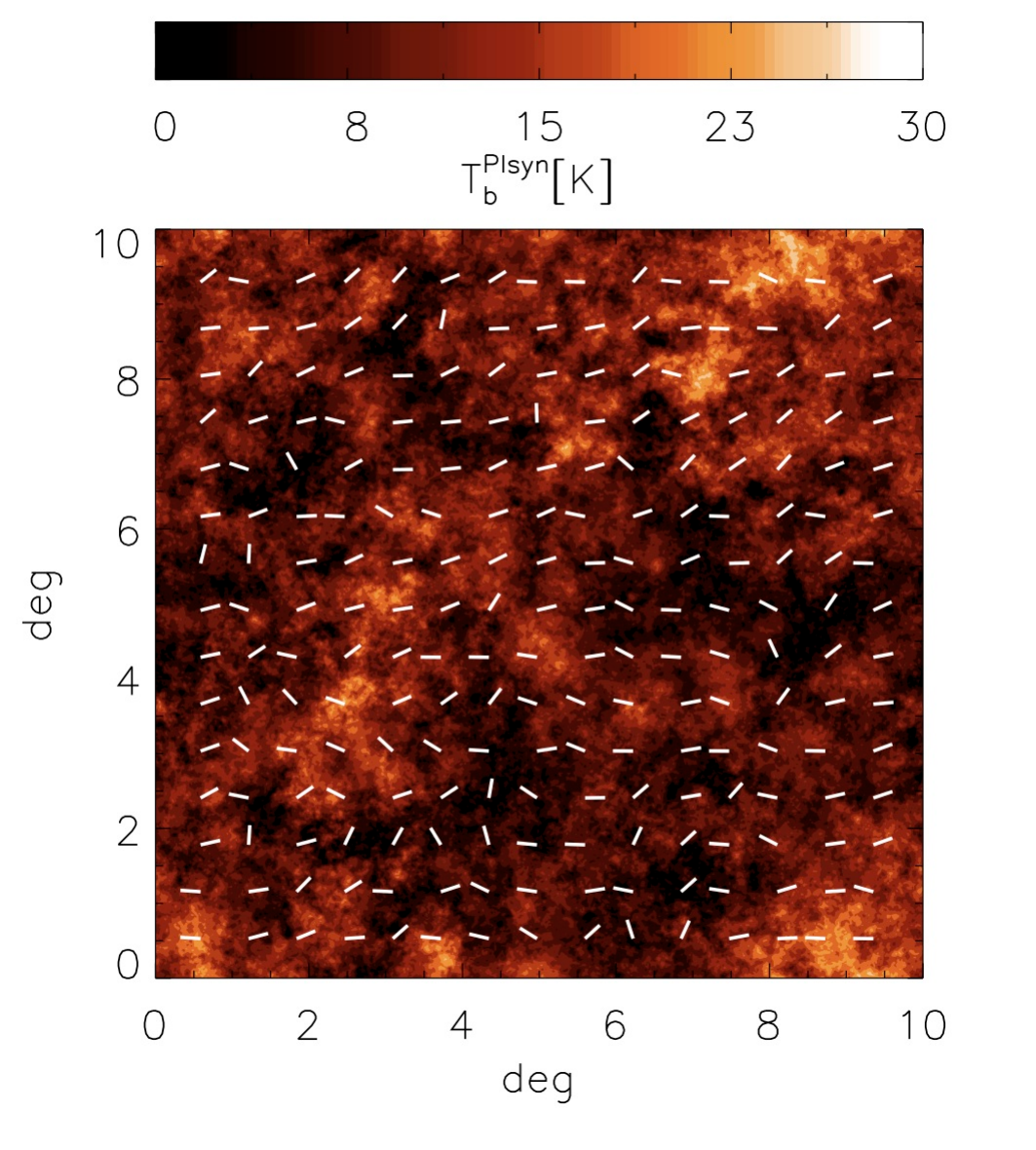}
\centering \includegraphics[width=.32\textwidth]{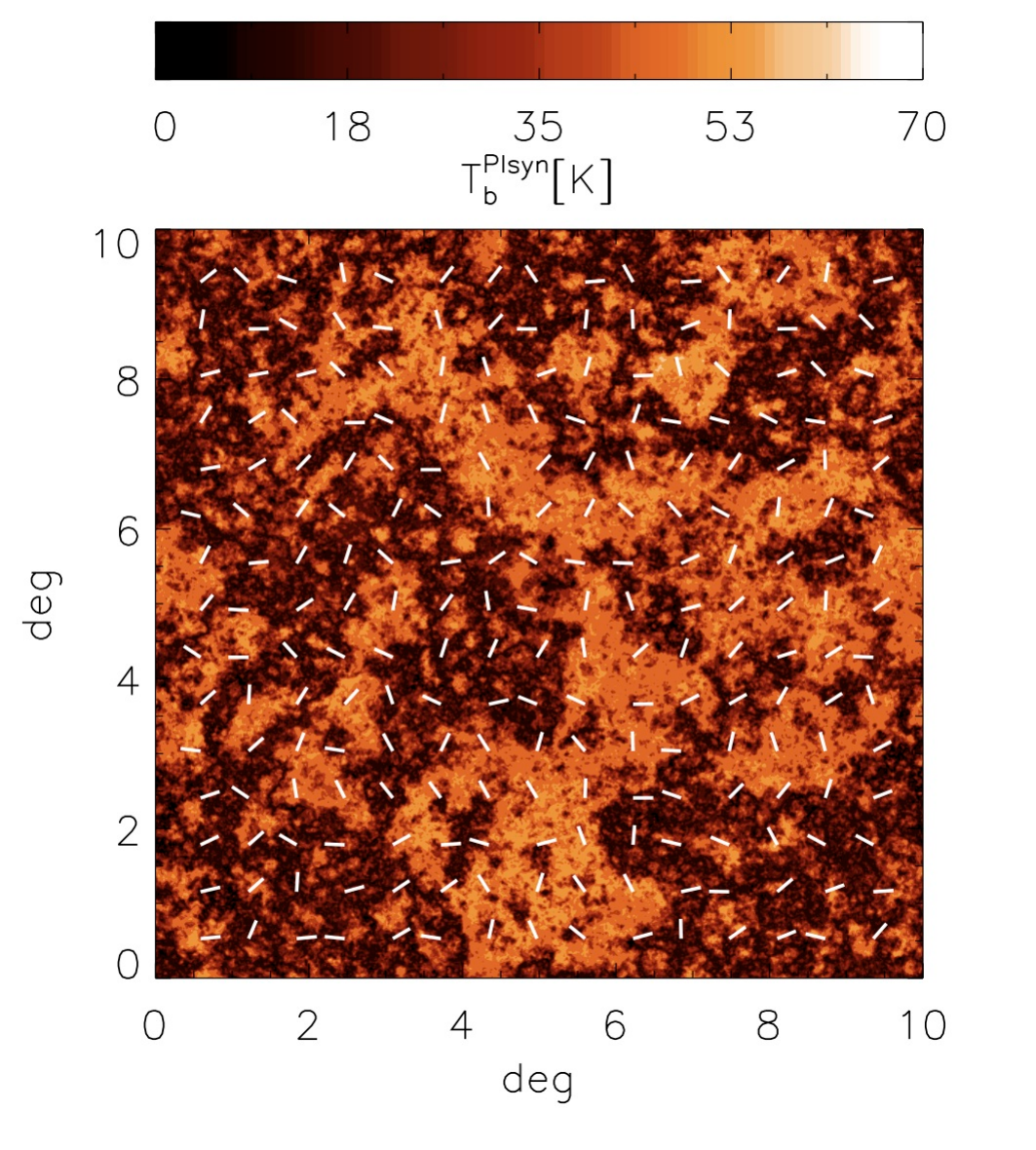}
\\ (a)\hspace{5.6cm}(b)\hspace{5.6cm}(c)
\caption{Simulated maps of the polarized intensity and polarization
angle (white lines) of the four different Galactic synchrotron
emission models (A, B and C from left to right). The angular size 
of the maps are $10^\circ\times10^\circ$, with $\sim 1~{\rm arcmin}$
resolution. The $mean$ and $rms$ values of the maps at $150~{\rm MHz}$ 
are given in the Table~\ref{tab:maps}.}
\label{fig:PImodels}
\end{figure*}

\begin{figure*}
\centering \includegraphics[width=.32\textwidth]{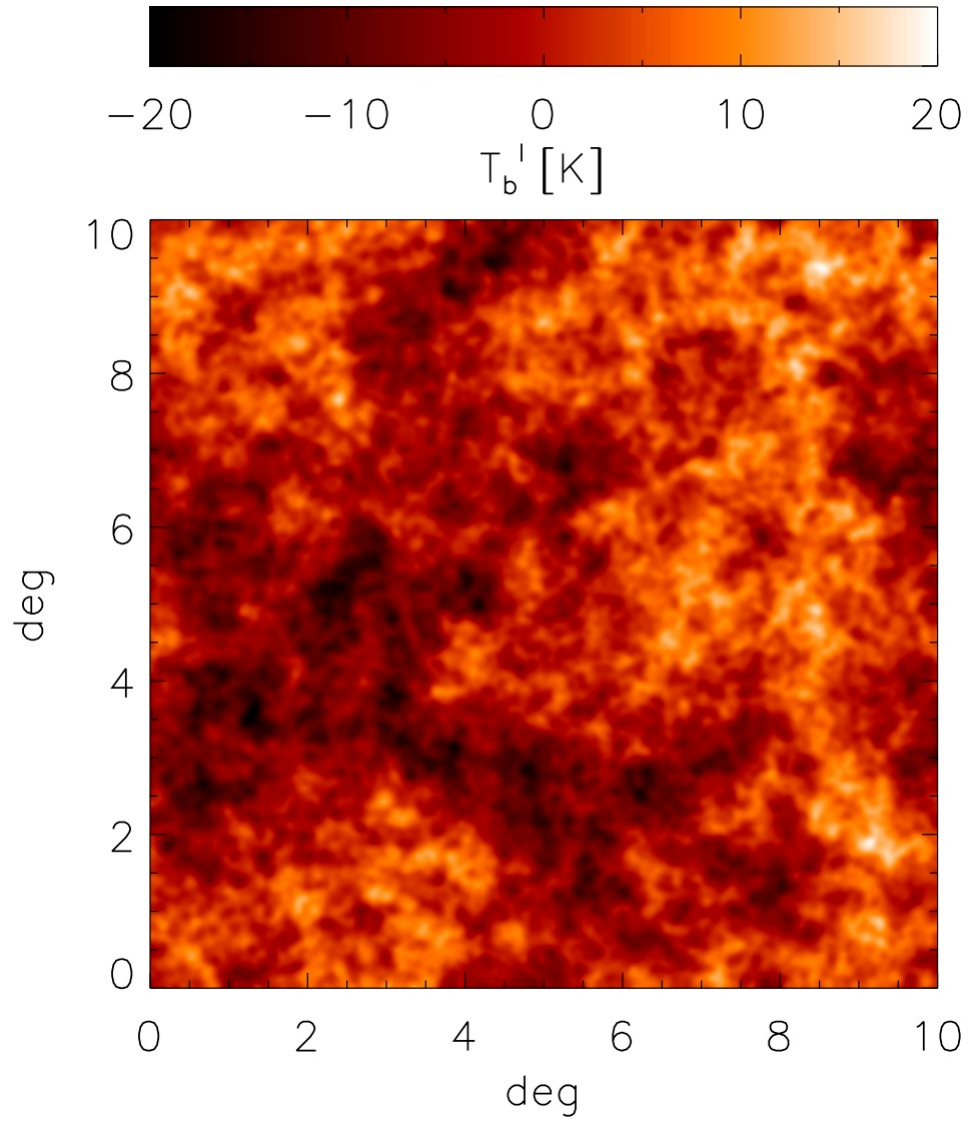}
\centering \includegraphics[width=.32\textwidth]{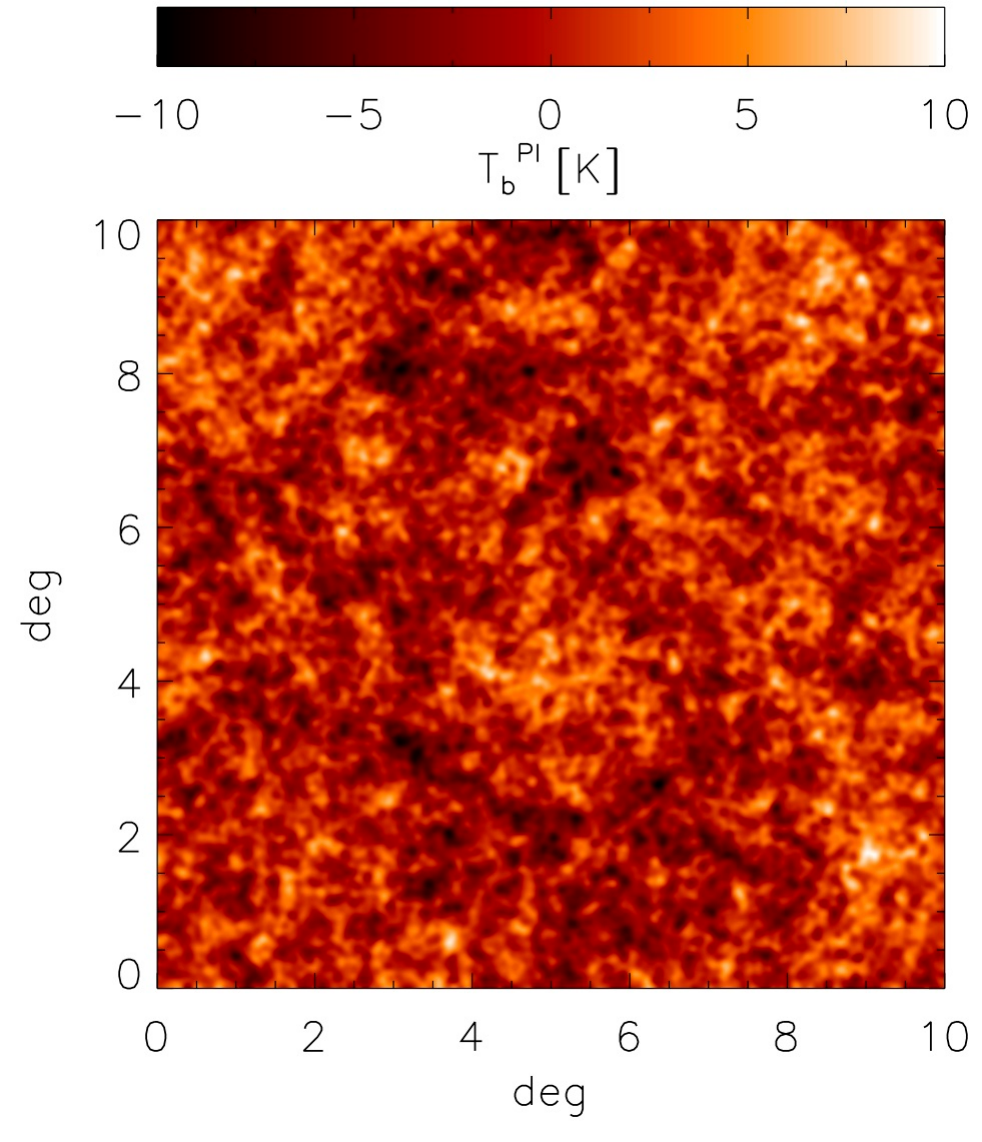}
\centering \includegraphics[width=.325\textwidth]{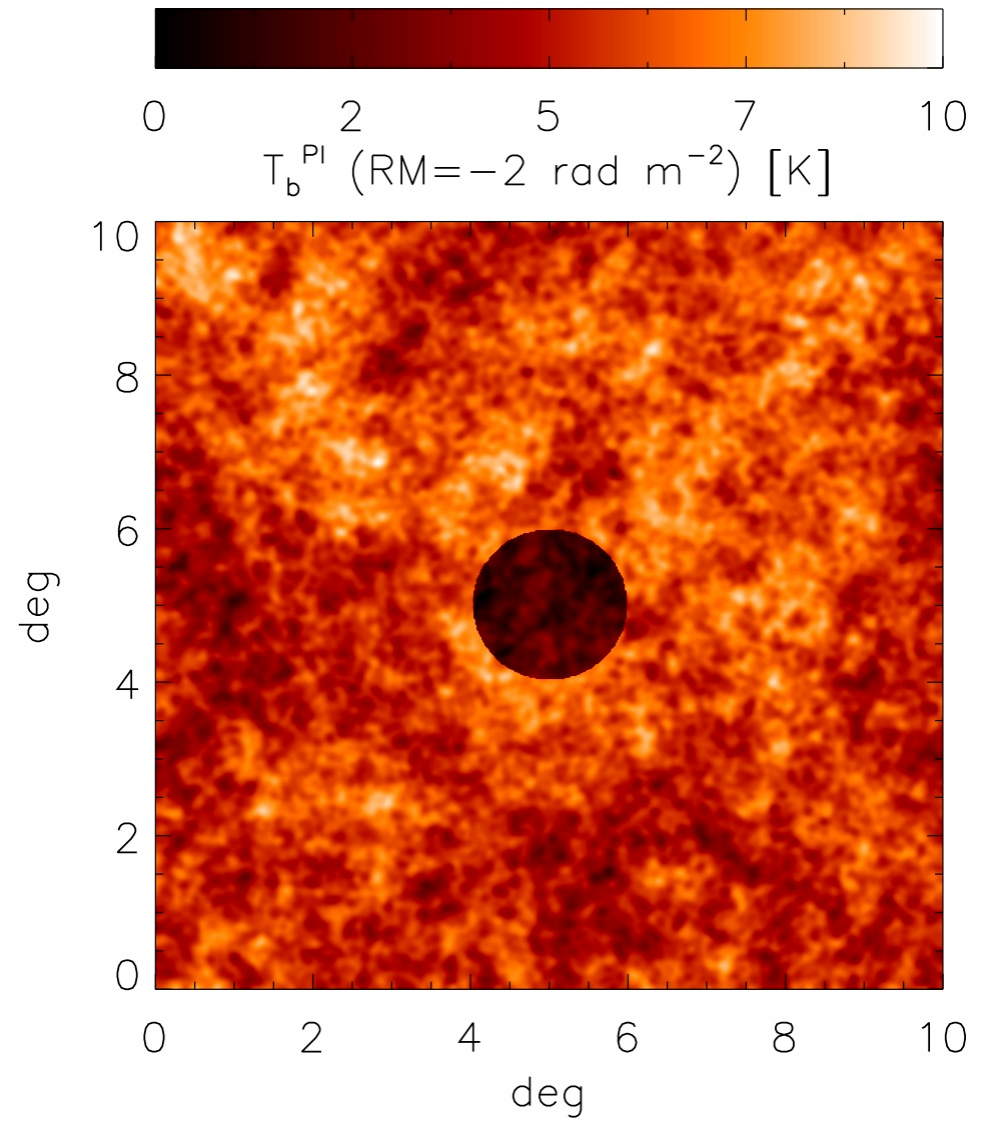}
\\ (a)\hspace{5.6cm}(b)\hspace{5.6cm}(c)
\caption{Simulated maps of the Fan region (model D) in total (Fig.~\ref{fig:Fan} a) and 
polarized intensity (Fig.~\ref{fig:Fan} b). An RM image obtained by applying rotation 
measure synthesis technique on the simulated data is shown in Fig.~\ref{fig:Fan} c.
}
\label{fig:Fan}
\end{figure*}

Further we explore morphological characteristics 
of the simulated polarized emission by comparing the image of
intrinsically polarized emission (Fig.~\ref{fig:synff}.b)
with the images of polarized emission of the first three
 models (see Fig.~\ref{fig:PImodels}).
Note that degree of polarized and depolarized emission is given in
Table~\ref{tab:maps}.
 
Model A assumes that there is no region in which the plasma
(thermal electron cloud) is mixed with CR electrons. Therefore,
polarization angle of synchrotron emission along the LOS are 
Faraday rotated by an equal amount (defined by Eq.~\ref{eq:FR}). 
Since there is no differential Faraday rotation, depolarization does
not occur and the polarized intensity of the synchrotron emission 
is unchanged (see Fig.~\ref{fig:PImodels}.a and Table~\ref{tab:maps}). 
Morphology of the polarization
angles over the map is determined only by the spatial $RM$ variations
of the plasma.

In order to test the above results, we correlate a polarized emission 
image with a total intensity, and polarization angles with the image of 
free-free emission. The cross correlation coefficient\footnote{The cross
correlation coefficient between two images ($a_{i,j}$
and $b_{i,j}$) with the same total number of pixels $n$ is defined as:
\begin{equation}\label{eq:c0}
C_0=\frac{1}{n-1}\sum_{i,j} \frac{(a_{i,j} - \bar{a})
(b_{i,j}-\bar{b})}{\sigma_{a}\sigma_{b}},
\end{equation}
where $\bar{a}$ ($\bar{b}$) is the mean and $\sigma_{a}$
($\sigma_{b}$) the standard deviation of the image $a$ ($b$). }
is $C=1$ in both cases.
Since the polarized emission s completely correlated with the total intensity,
and the free-free emission, which traces thermal plasma, with polarization angles, 
we conclude that
our algorithm is preforming as expected. The spatial structures of 
polarized emission are unchanged, while the structures of polarization angle
follow thermal plasma.

In B and C models there are regions where both CR
and thermal electrons are mixed. In those regions the polarization angle of
the synchrotron radiation is then differentially Faraday rotated along
the LOS. As a result, the polarized synchrotron radiation is quenched and
the level of the polarized emission is weaker than intrinsic polarization emission
(see Table~\ref{tab:maps}).
In addition to depolarization produced by the mixed regions, in model C 
depolarization is also produced by variations of the Galactic magnetic field 
along the LOS.

As for model A, we calculate correlations between the
total and polarized images. In both models (B \& C) we find much weaker correlation 
($C_{B,C}=0.3,0.1$) than in model A. The weaker correlation results from differential 
Faraday rotation along the LOS, which changes the morphology of polarized 
emission both in intensity and angle.  To illustrate this we correlate the polarized
synchrotron emission with the free-free emission. We find 
anti-correlation in both models ($C_{B,C}=-0.7,-0.5$). The intensity of the polarized emission is 
more quenched in the regions that show stronger free-free emission (regions of denser plasma).
Moreover, model C shows weaker (anti)correlations than model B, due to Faraday 
structures produced by magnetic field.

\begin{table}
  \centering
  \caption{The $mean$ and $rms$ value of the maps shown in the
  Fig.~\ref{fig:synff}~\&~\ref{fig:PImodels}. All the values are given
  in kelvin. For completeness, degree of polarized ($PI/T$) and
  depolarized ($dep.$) emission is calculated.}
  \begin{tabular}{ccccccc}
    \hline & $I_{syn}$ & $I_{ff}$ & $iPI_{syn}$ & $PI_A$ & $PI_B$ &
    $PI_C$\\ \hline $\rm mean$ & 142 & 1.5 & 98 & 98 & 10 & 11
    \\ $\rm rms$ & 3 & 0.1 & 2 & 2 & 5 & 5 \\ $PI/I$ & -  & -  & 69\% &
    69\% & 7\% & 8\% \\ $dep.$ & - & - & 0\% & 0\% & 90\% &
    88\% \\\hline
  \end{tabular}\label{tab:maps}
\end{table}

\begin{figure*}
\centering \includegraphics[width=.95\textwidth]{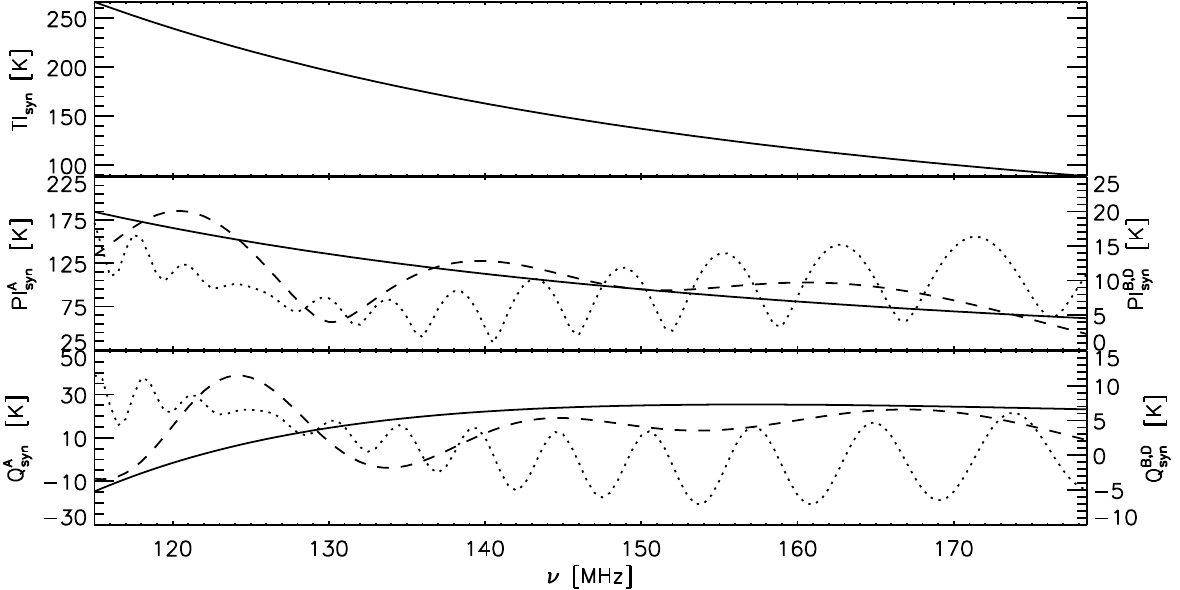}
\caption{A random line through synchrotron total intensity
($TI_{syn}$) and polarized intensity ($PI_{syn}, Q_{syn}$) frequency
data cubes. The \textit{solid lines} are for the Galactic
\textsc{model A}, \textit{dashed lines} for \textsc{model B} and
\textit{dotted lines} for \textsc{model D}. $TI_{syn}$ is the same in 
all four models, while $U_{syn}$ is not presented since it is 
similar to $Q_{syn}$. Note the polarized structures along the 
frequency direction. An improper polarization calibration of the
instrument could cause a leakage of these structures to the 
total intensity and severely contaminate the EoR signal.}
\label{fig:los}
\end{figure*}

The last aspect of the algorithm, which needs to be verified, is
the frequency behavior of the simulated Galactic emission. It is expected 
that the synchrotron emission shows roughly a power law along the
frequency (see Eq.~\ref{eq:syn}). By examining Fig.~\ref{fig:los}, one
can see that simulated synchrotron emission indeed shows a power law
behavior in total intensity. However, in polarized intensity only model A
shows a power law behavior. All the other models show characteristic 
fluctuations caused by differential Faraday rotation \citep{burn66}. Note that fluctuations 
show richer structures, as in model C, if the differential Faraday rotation is more 
prominent.

The calculated brightness temperature spectral
index, $\beta_{syn}$, variations of the simulated Galactic 
synchrotron emission are given in the Fig.~\ref{fig:synff}.c. The
variations of $\beta_{syn}$ are estimated by fitting the data with a linear 
functions along the frequency in logarithmic scale. The $mean$ value 
of $\beta_{syn}$ is -2.50, which is in a good agreement
with the expected theoretical value $\beta_{p=2}=-(p+3)/2=-2.55$. 
A slight difference between the two is caused by the 3D spatial variations 
of the spectral index $p$.

Based on discussion and tests of the algorithm presented in this
subsection, we can conclude that our algorithm is simulating Galactic emission
as expected. In the following subsection, we will compare the simulated
data cubes with the observations.

\subsection{Comparison with observations}
Input parameters of the simulation (properties of the CR electrons, thermal 
electrons, and Galactic magnetic field, see subsections~\ref{sec:CRe}, \ref{sec:MF} 
\& \ref{sec:Te}) are chosen in such way that the global properties (boundary conditions) 
of the simulated maps, e.g. morphology and frequency behavior of the Galactic emission, 
are in agreement with the observations overviewed in Sec. 3:
\begin{itemize}
\item the mean value of the simulated synchrotron emission in total intensity, $I_{syn}=142~{\rm K}$ 
(see Table~\ref{tab:maps}) is comparable to observed emission at 150~{\rm MHz} \citep[in the cold regions of the Galaxy at the high galactic latitudes, ][]{landecker70} or to the values obtained by extrapolation from the higher frequencies, i.e., $\sim150~{\rm K}$;   
\item the mean and variations of the brightness temperature spectral index of the simulated maps, 
$\beta_{syn}=-2.50$ (see Subsection~\ref{sec:dta}) and $\sigma=0.1$, are in agreement with the observed
values, $\beta^{obs}_{syn}=-2.55$ and $\sigma^{obs}_{syn}\approx0.1$ \citep[e.g.][and references therein]{rogers08};
\item the spatial structures of the simulated Galactic emission in total intensity are morphologically similar to the observed emission around $150~{\rm MHz}$ \citep{bernardi09a,bernardi10} and $\sim300-400~{\rm MHz}$ \citep[e.g.][and references in Sec.~\ref{sec:obs}]{haslam82, debruyn98}, i.e., power law type of the power spectrum with a negative power law index (more power on the large scales);
\item the simulated polarized emission (e.g. model B, C \& D) shows small-scale structures, which have no counter parts in the total intensity, e.g. like a number of observations with the WSRT (see Sec.~\ref{sec:obs});
\item the simulated distribution of the thermal electrons reproduce the typical observed $EM\sim 10~{\rm cm^{-6}pc}$ at the high Galactic latitudes \citep{berkhuijsen06}.
\item the simulated Galactic emission (see Fig.~\ref{fig:los}) shows characteristic power law behavior along the frequency in total intensity, while in polarized intensity can reproduce the characteristic polarization functions \citep{burn66}.
\end{itemize}

We also quantitatively compare the simulated emission in the Fan region (our model D)
with the observations of this region at $150~{\rm MHz}$ \citep{bernardi09a}:
\begin{itemize}
\item simulated fluctuations have an rms of $14~{\rm K}$ in total intensity and rms of
$7~{\rm K}$ in polarized intensity (these values are the same as the observed ones);
\item the power law indices of the power spectrum obtained from the simulation are 
in agreement with the observations: $-2.2$ in total intensity, and $-1.6$ in polarized intensity;
\item simulated maps (see Fig.~\ref{fig:Fan}) show similar morphological structures (their statistical distribution of power) as the observed 
ones \citep[fig 5., 8. \& 10. in][]{bernardi09a};
\item a simulated rotation measure cube\footnote{The rotation measure cube is obtained using the rotation measure synthesis method \citep{brentjens05}.} shows similar morphological structures 
as the observed ones, i.e., at $-2~{\rm rad~m^{-2}}$ there is a hole in the emission,
a bubble with a diameter of $\sim2^\circ$.
\end{itemize}

Here we emphasize that we have also successfully tested our algorithm at the higher frequencies 
($\sim350~{\rm MHz}$) by simulating Galactic emission towards the cluster Abel 2255. The simulated maps have been compared with the observations obtained by R.F. Pizzo et al. (\textit{in prep., private communication}). Details of these simulations  will be presented
in a separate paper (Jeli{\'c} et al., \textit{in prep.}).

To summarize,  simulated maps show all observed characteristics of the Galactic emission, 
e.g. presence of the structures at different scales, spatial and frequency variations of the brightness 
temperature and its spectral index, complex Faraday structures, and depolarization. Based on these results, we conclude that our model is able to simulate realistic maps of the Galactic emission both in total and polarized intensity, and can be used as an realistic foreground model in simulations of the EoR experiments.

\section{LOFAR-EoR simulation pipeline}\label{sec:lofar-eor}
The LOFAR-EoR project relies on a detailed understanding of
astrophysical and non-astrophysical contaminations that can contaminate the
EoR signal: the Galactic and extragalactic foregrounds, ionosphere,
instrumental effects and systematics. In order to study these
components and their influence on the detection of the EoR signal, a
LOFAR-EoR simulation pipeline is being developed by the LOFAR-EoR
team. The pipeline consists of the three main modules: the EoR
signal \citep[based on simulations described in][]{thomas09}, 
the foregrounds \citep[based on this paper and ][]{jelic08} and the 
instrumental response \citep[described on][]{panos09}.

In this paper we use the LOFAR-EoR pipeline to illustrate the
need for excellent calibration of the instrument in order
to reliably detect the EoR signal (see
Sec.~\ref{sec:calibration}). The following two subsections present a
brief overview of the EoR signal and instrumental response modules.

\subsection{EoR signal}
The predicted differential brightness temperature of the cosmological
21~cm signal with the CMB as the background is given by
\citep{field58, field59, ciardi03}:
\begin{eqnarray}
\delta T_{\rm b}&=&26~\mathrm{mK}~x_{\rm HI}(1+\delta)
 \left(1-\frac{T_{\rm CMB}}{T_{\rm s}}\right) \left(\frac{\Omega_{\rm
 b} h^2}{0.02}\right)\nonumber\\ &&\left[\left(\frac{1 +
 z}{10}\right)\left(\frac{0.3}{\Omega_{\rm m}}\right)\right]^{1/2}.
\label{eq:tbright0}
\end{eqnarray}
Here $T_{\rm s}$ is the spin temperature, $x_{\rm HI}$ is the neutral
hydrogen fraction, $\delta$ is the matter density contrast and
$h=H_{0}/(100~{\rm km~s^{-1} Mpc^{-1}})$. Throughout we assume
 $\Lambda$CDM-cosmology with WMAP3 parameters
\citep{spergel07}: $h=0.73$, $\Omega_{\rm b}=0.0418$,
$\Omega_{\rm m}=0.238$ and $\Omega_\Lambda=0.762$. In addition we 
assume that $T_s \gg T_{\rm CMB}$, which is assumed in most of the 
current simulations.

\begin{figure*}
\centering \includegraphics[width=1.\textwidth]{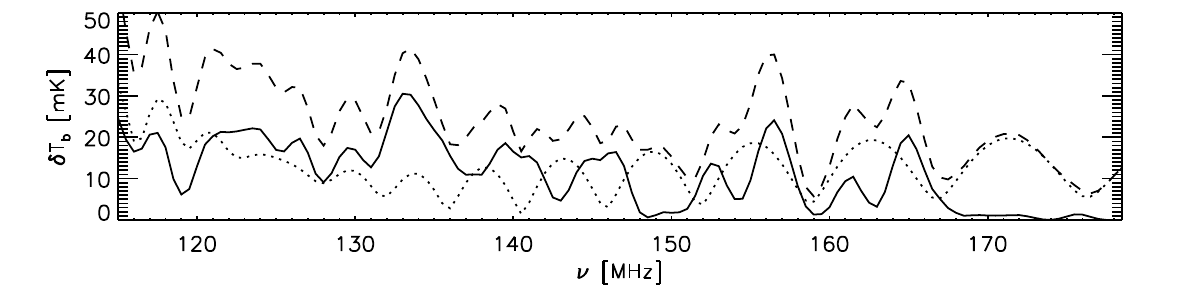}
\caption{A random line of sight through a simulated 21~cm data cube for the
 `Stars' patchy reionization history model (\textit{solid line}). \textit{Dotted line} 
 shows the `leakage' of the polarized Galactic emission to the total intensity and
 \textit{dashed line} is a sum of the two. We assume 0.15\%  residual `leakage'
 and we use model D as an example of the Galactic emission. The angular 
 and frequency resolution of the data match that of the LOFAR telescope.}
\label{fig:losEOR}
\end{figure*}

The cosmological 21~cm maps ($\delta T_b$) are simulated using the
\textsc{bears} algorithm \citep{thomas09}. \textsc{bears} is a fast
algorithm to simulate the underlying cosmological 21~cm signal from
the EoR. It is implemented using an N-body/SPH simulation in
conjunction with a 1-D radiative transfer code under the assumption of
spherical symmetry of the ionized bubbles. The basic steps of the
algorithm are as follows: first, a catalogue of 1D ionization profiles
of all atomic hydrogen and helium species and the temperature profile
that surround the source is calculated for different types of ionizing
sources with varying masses, luminosities at different
redshifts. Subsequently, photon rates emanating from dark matter
haloes, identified in the N-body simulation, are calculated
semi--analytically. Finally, given the spectrum, luminosity and the
density around the source, a spherical ionization bubble is embedded
around the source, whose radial profile is selected from the catalogue
as generated above. For more details we refer to \citet{thomas09}.

For the purpose of this paper we use the $\delta T_b$ data cube (2D
slices along the frequency/redshift direction) of the cosmological
21~cm signal for the `Stars' patchy reionization model
\citep[see][]{thomas09}. The data cube consists of 850 slices in
the frequency range from $115~{\rm MHz}$ to $200~{\rm MHz}$ with
$0.1~{\rm MHz}$ step (corresponding to redshift between 6 and 11.5).
Slices have a size of $100~\rm{h^{-1}}$ comoving Mpc and are defined
on a $512^2$ grid. An example of a random line of sight through 
simulated 21~cm data cube, with angular and frequency resolution
matching that of LOFAR, is shown in Fig.~\ref{fig:losEOR}.

\subsection{Instrumental response}
In order to produce the dirty maps of the diffuse emission, we calculate
the 2D Fourier transform of the data for each correlation on a fine
grid of $1.2~{\rm arcmin}$. We assume that there are 24 stations in the 
LOFAR array that
are used for the observations. We then use a bilinear interpolation to
estimate the values of the visibilities at the $uvw$ points that correspond 
to the points sampled by the interferometer pairs of the core. This is done
for 4hrs of synthesis, 10 sec integration and for the whole frequency range
between 115 MHz and 180 MHz, using a step of 0.5 MHz. The above 
procedure is implemented as a parallel algorithm in the CHOPCHOP
pipeline (see Labropoulos et al, \textit{submitted}).  In order
to sample the large structure of the foregrounds at scales between 5 and 
10 degrees we need interferometer spacing between 6.5 and 13 meters. 
Thus the PSF acts as a high-pass spatial filter. 

\begin{figure*}
\centering \includegraphics[width=.33\textwidth]{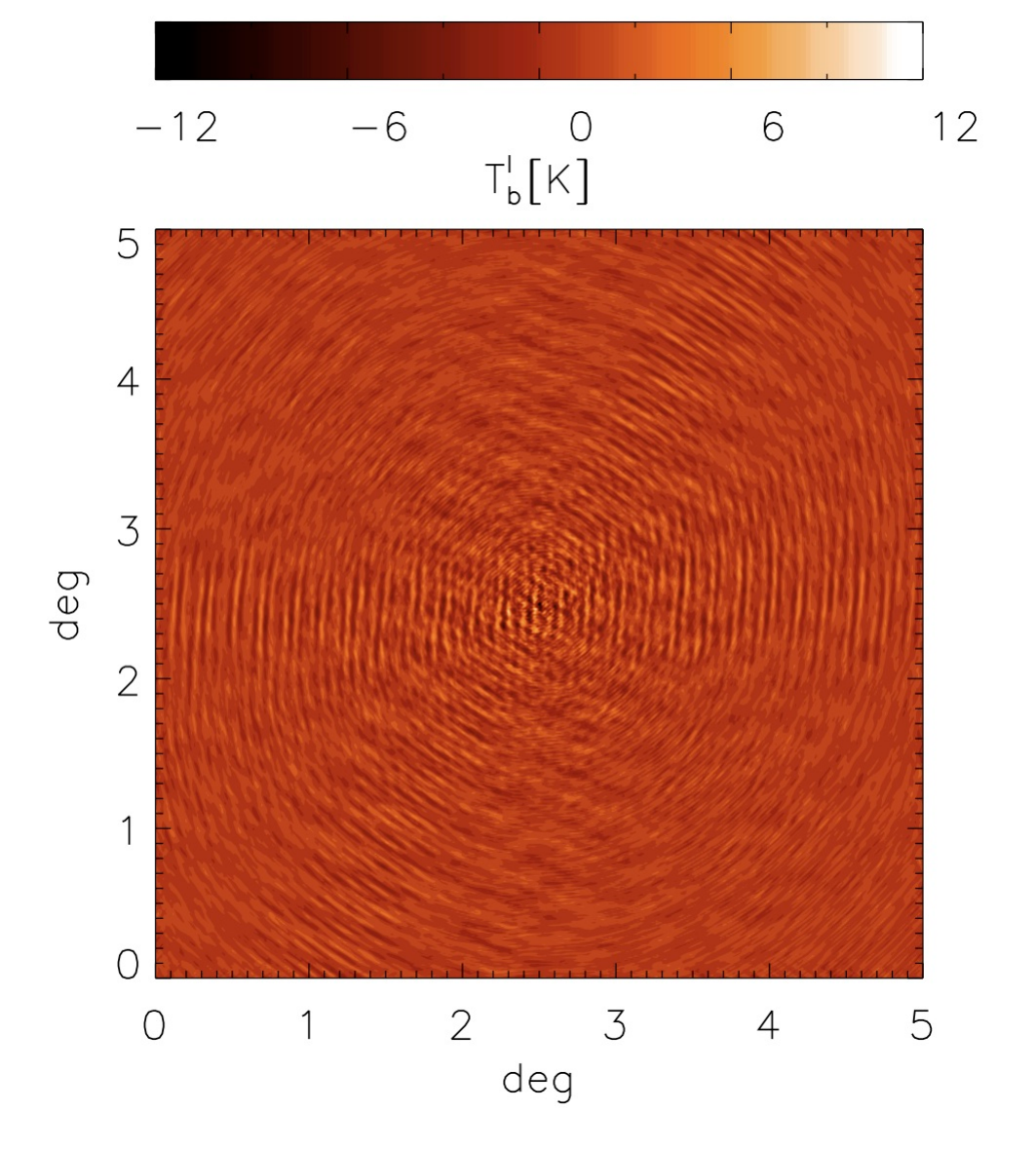}
\centering \includegraphics[width=.33\textwidth]{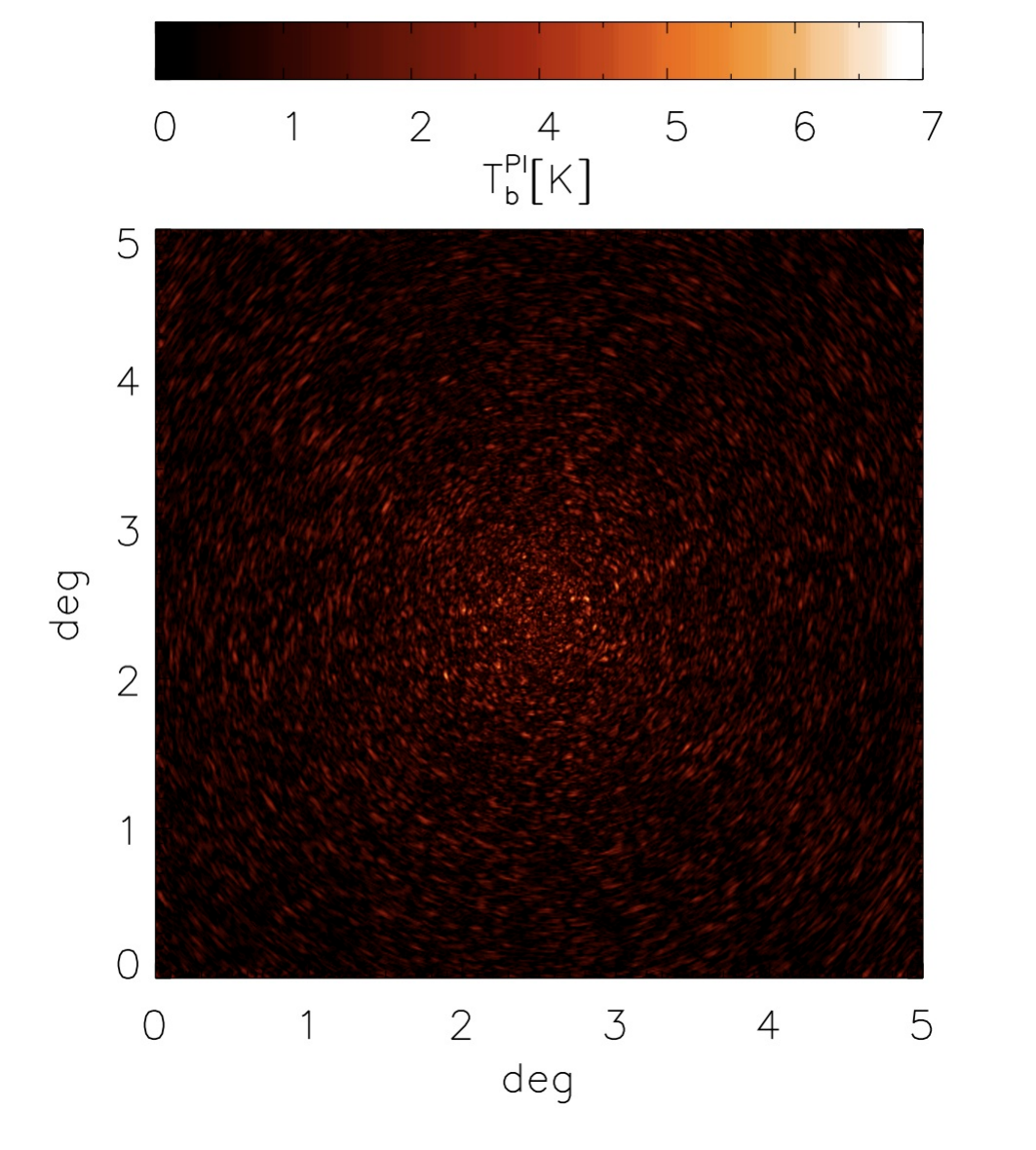}
\centering \includegraphics[width=.33\textwidth]{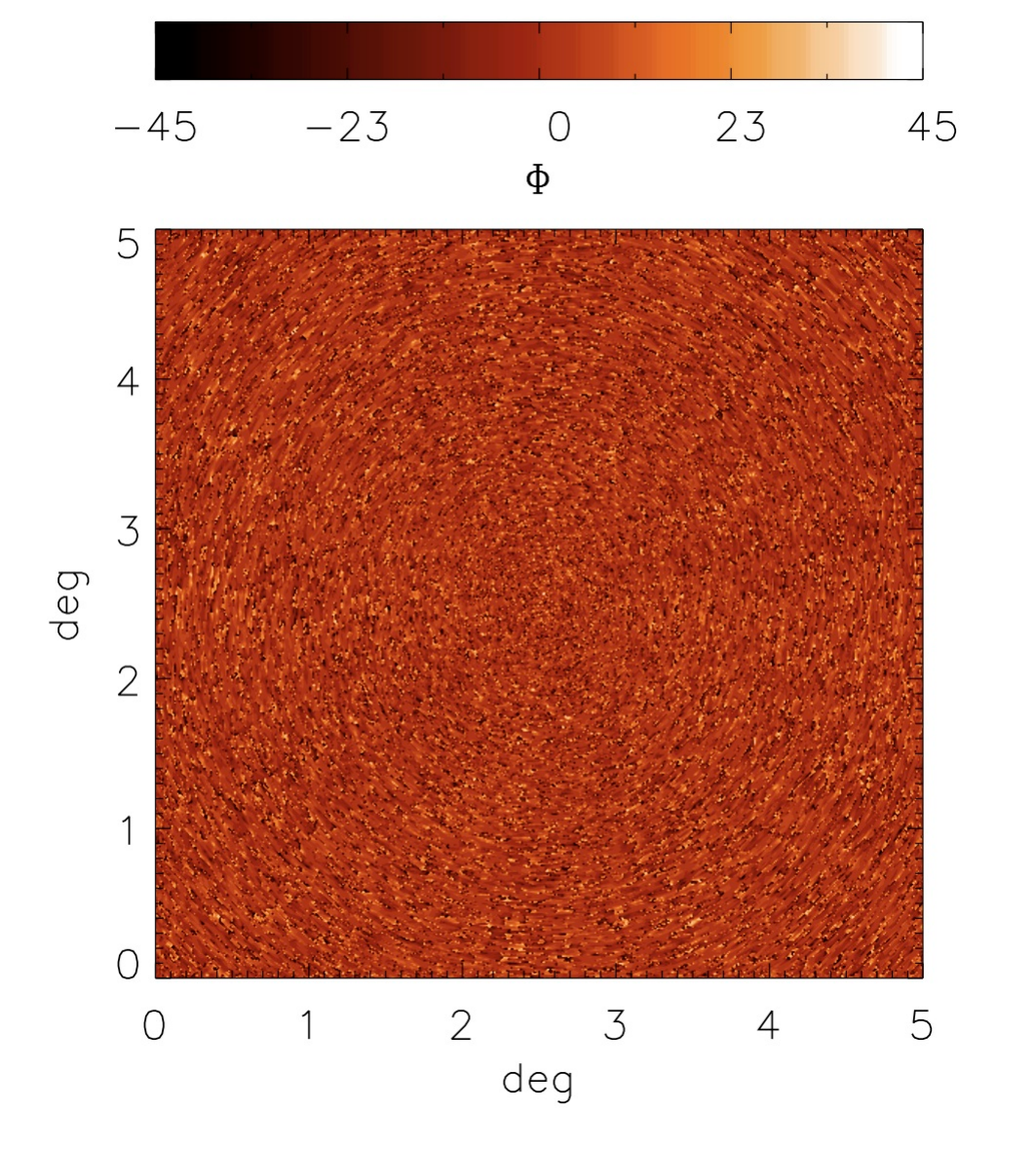}
(a)\hspace{5.6cm}(b)\hspace{5.6cm}(c)
\caption{`Dirty' maps of the simulated Galactic synchrotron emission 
(\textsc{model B}) observed with the core stations of the LOFAR 
telescope. The total and polarized intensity maps are shown in
Fig.~\ref{fig:dmap}.a and Fig.~\ref{fig:dmap}.b, while the polarization angle 
is presented in Fig.~\ref{fig:dmap}.c. The images are simulated at $138~MHz$}.
\label{fig:dmap}
\end{figure*}

Figure \ref{fig:dmap} shows `dirty' maps of the simulated Galactic 
synchrotron emission (\textsc{model B}) observed with the core 
stations of the LOFAR telescope at $138~{\rm MHz}$. The total and polarized
intensity maps are shown in Fig.~\ref{fig:dmap}.a~\&~b,
while the polarization angle is presented in Fig.~\ref{fig:dmap}.c. Note 
that the large scale structures of the emission are missing as the smallest 
baseline length is approximately $50~{\rm m}$.

\section{Consequences for 21-cm Reionization Detection Experiments}\label{sec:calibration}
One of the major challenges of the EoR experiments is the extraction
of the EoR signal from the astrophysical foregrounds. The 
extraction is usually formed in total intensity along the 
frequency direction due to the following characteristics:
\begin{enumerate}
\item the cosmological 21~cm signal is essentially unpolarized and fluctuates
along the frequency direction (see Fig.~\ref{fig:losEOR})
\item the foregrounds are smooth along the frequency direction 
in total intensity and should only show fluctuations in polarized intensity 
(see Fig.~\ref{fig:los}, an example of the Galactic emission that is a 
dominant foreground component). 
\end{enumerate}
Thus, the fluctuating EoR signal can be extracted from the foregrounds 
by fitting the smooth component of the foregrounds out \citep[e.g. see fig. 12 in][]{jelic08}.

All current EoR radio interferometric arrays have an
instrumentally polarized response, which needs to be calibrated.
If the calibration is  imperfect, some part of the polarized signal is
transferred into a total intensity and vice versa (hereafter 
`leakages').  As a result, the extraction of the EoR signal is 
more demanding.

Moreover, the polarized signal could have similar 
frequency fluctuations as the cosmological signal and as such 
could possibly severely contaminate it. Thus, to reliably detect the
cosmological signal it is essential to minimize the `leakages' and to
observe in the regions with the weak polarized foreground emission.
We illustrate this through an example for the LOFAR telescope,
but the problem is common to all current and planned EoR radio arrays.

\subsection{Levels of `leakage'}
The `leakages' of the total and polarized signal are produced by two 
effects: the geometry of the LOFAR array and  the cross-talk between the 
two dipoles in one LOFAR antenna. The cross-talk, a leakage in the 
electronics that can cause the power from one dipole to be
detected with other, is small  compared to the geometric effects 
(Labropoulos et al, \textit{submitted}) and we will ignore it for 
the purpose of this paper. 

The geometry of the LOFAR telescope is such that the array antennae 
are fixed to the ground. Therefore, the sources are tracked only by 
beam-forming and not by steering the antennae mechanically towards 
 the desired direction. This implies that, depending on the position 
of the source on the sky, a non-orthogonal (except at the zenith) projection 
of the two orthogonal dipoles is visible by the source. This projection 
further changes as the 
source is tracked over time. Thus, the observed Stokes brightness of 
the source, ${\bf S}_{obs}$, is given by \citep{carozzi09}:
\begin{equation}
{\bf S}_{obs}={\bf M}{\bf S},
\end{equation}
where ${\bf M}$ is a Mueller matrix that quantifies the distortions of a 
true source brightness  ${\bf S}={\bf (}I,Q,U,V{\bf)}$ based on above 
geometry-projection effect. The Mueller matrix is defined as:
\begin{equation}\label{eq:leak}	
{\bf M}= \left ( 
	\begin{array}{cccc}
  	\frac{1}{2}(1+n^2) &	 -\frac{m^2}{2} + \frac{(1+m^2)l^2}{2(1-m^2)} & -\frac{lmn}{1-m^2} & 0\\
  	\frac{1}{2}(l^2-m^2) & 1-\frac{m^2}{2} - \frac{(1+m^2)l^2}{2(1-m^2)} & \frac{lmn}{1-m^2} & 0\\
  	-lm & -lm & n & 0 \\
	0 & 0 & 0 & n \\
	\end{array} 
\right )
\end{equation}
with the assumption of  a coplanar array. Note that $(l,m,n)$ are direction cosines
and that the level of geometry-projection `leakage' therefore varies across the map.

The calculated leakages, for the LOFAR telescope
observing at $138~{\rm MHz}$ a $5^\circ \times5^\circ$ patch of the 
sky around the zenith, vary between 0.1\% -- 0.7\% across the image. 
For the sky model we use the total and polarized intensity maps of 
Galactic emission (model C). Further, an instant imaging is assumed, 
i.e. the sky is not tracked over time. Note that the leakages are tiny around the 
center of the image, but they increase towards the edges. 

The same calculation we repeat for an patch of a sky at  
$45^\circ$ altitude. The leakages are now much larger, e.g.
for the center of the image the leakage to the total intensity is
around $2\%$, but can reach $20\%$ towards the edges of the
image. 

Once the tracking of the sources is taken into account,
the leakages become even more significant towards
the horizon and on the axes parallel to the dipoles, i.e. 
varying between 0.1\%--100\% across the sky (for
details we refer to Labropoulos et al, \textit{submitted}).

However to be as realistic as possible, we need to take 
into account that for calibration of the instrument one can 
use a model of the beam. Using the beam model, one
can correct for the polarized instrumental response and 
decrease the level of `leakages'. Thus, in further discussion
 we refer only to the residual `leakages'. Note that 
 for the purpose of this paper we assume that the 
 residual `leakages' are 1\%--5\% over the whole field of view.

\subsection{`Leakage' of the polarized foreground}
Here we would like to point out  that the residual
`leakages' caused by the geometry-projection effect are
significant in terms of EoR signal detection. If these `leakages'
are not taken properly into account during the calibration of the instrument, the 
polarized Galactic emission could creep into total intensity
signal and severely contaminate the EoR signal. This is illustrated in 
the Fig.~\ref{fig:losEOR}.

Figure~\ref{fig:losEOR} shows a random line of sight through
 a simulated 21 cm data cube for the `Stars' patchy reionization history 
 model (solid line). As noted before, the EoR signal 
 shows fluctuations in total intensity. Dotted line shows the `leakage'
 of the polarized Galactic emission in the total intensity and dashed line is 
 a sum of the two.
 
Since we have chosen  the Galactic emission model with the 
 differential Faraday rotation (model C), the `leakage' of the Galactic 
 emission shows structures along the frequency direction similar to 
 the EoR signal. Note that the Galactic polarized emission
is assumed to be $\sim 1~{\rm K}$ (in this case our model C), while the level of
   the residual `leakage' is assumed to be $1.5\%$.  The
  `leaked' Galactic polarized emission is then of the same order as the 
  EoR signal ($\sim 15~{\rm mK}$).
 
In other words, the right combination of the Galactic polarized 
emission and an inaccurate calibration can result in structures 
in total intensity that have similar characteristics of the EoR signal. 
Without knowing the exact characteristics of the Galactic polarized 
emission in that region it would be impossible to extract the EoR signal.
Therefore, the observational windows for the EoR experiments need
be in regions of the Galaxy that have very weak or no polarization.
Assuming the same residual `leakage' of 1.5\%, the Galactic polarized 
emission should not be stronger than  $\sim 0.1~{\rm K}$. 
The 'leaked' foreground contamination would then be at least an order of magnitude lower than
the EoR signal, which is acceptable for the EoR detection.

However, note that this upper limit on the polarized contamination
highly depends on morphology of the Galactic emission. For example, 
if the Galactic polarized emission does not show structures along 
the frequency direction (like in our model A), then any `leaked' 
contamination can be fitted out in the same way as the foregrounds in 
total intensity. Moreover, if the Galactic polarized
emission shows significant structures on scales smaller then resolution of
the array, beam depolarization will lower the level of the observed polarized 
foregrounds and therefore lower the `leaked' contamination as well.

Other possible ways of eliminating a `leaked' polarized foreground 
from the data are: (i) identifying the polarized emission in rotation measure space
using a Faraday rotation measure synthesis \citep{brentjens05}; (ii) using  
polarization surveys obtained by a different radio telescope and with angular 
resolution that is preferably higher than the one of the 
EoR experiments; and (iii) observing the EoR in a multiple regions of the sky.

Note that if someone uses a different polarization data as an EoR foreground template,
an instrumental beam depolarization should be take into account. At the same time the advantage
is that radio telescopes, which are used for the polarization surveys, suffer from a different
systematics (including `leakages') and therefore comparison of the data 
observed with a different radio telescopes helps in analysis.

The observations of the EoR signal in multiple windows have advantage of allowing
us to cross-correlate the data and increase the significance of the EoR detection.  Since 
the Galactic emission varies across the sky, each observational window will have its own
characteristic foreground emission and `leakages'. Residuals after foreground removal
should not correlate between different observational windows, while the EoR properties
should be similar.

\section{Summary and conclusions}\label{sec:so}
This paper presents Galactic foreground simulations used as
templates for the LOFAR-EoR testing pipeline. The simulations provide
 maps of the Galactic free-free emission and the Galactic synchrotron 
 emission both in
total and polarized intensity. The maps are $10^\circ\times10^\circ$
in size, with $\sim1~{\rm arcmin}$ resolution and cover the frequency
range between 115 and 180 MHz pertaining to the LOFAR-EoR experiment.
The code however is flexible as can provide simulation over any scale
with any spatial and frequency resolution.

The Galactic emission is calculated from a 3D distribution of 
cosmic ray and thermal electrons, and the Galactic magnetic field.
The model assumes two magnetic field components: regular and
random. The latter magnetic field and the thermal electron density are
simulated as Gaussian random fields with power law power spectra. 
In addition, the spatial variations of the energy spectral
index $p$ of the cosmic ray electrons are introduced to mimic the
observed fluctuations of the brightness temperature spectral index
$\beta$. Note that all parameters of the simulation can be tuned to
any desired value and this allows to explore the whole parameter space.

The total and polarized Galactic maps are obtained for four
different models of Galactic emission (see
Fig.~\ref{fig:synff}~\&~\ref{fig:PImodels}).  The first assumes that
synchrotron and free-free emitters are spatially separated, such that
thermal plasma acts as a ``Faraday screen''. The amplitude of the
polarized emission is unchanged, while the polarization angles 
Faraday rotate. Other three simulation have regions where both
types of emitters are mixed in different ways. The synchrotron
emission is differentially Faraday rotated and depolarization
occurs (see Table~\ref{tab:maps}).

The main result of our simulations is that we are able to
produce realistic Galactic polarized emission that is comparable to observations, 
i.e., presence of the structures at different scales, spatial and frequency 
variations of the brightness temperature and its spectral index, complex 
Faraday structures, and depolarization. The importance of this result
comes from the fact that the planned EoR radio arrays have a polarized
response and the extraction of the EoR signal from the foregrounds is
usually performed along the frequency direction. The
Galactic foreground is a smooth function of frequency in a 
total intensity and it can show fluctuations in polarized intensity.
The EoR signal is expected to be unpolarized and to show fluctuations
along the frequency direction. Therefore, an imperfect calibration 
of the instrumental polarized response can transfer a fraction 
of the polarized signal into a total intensity. As a result, the leaked polarized 
emission can mimic the cosmological signal and make its extraction almost
impossible (see Fig.~\ref{fig:losEOR}).
 
Based on our simulations, we conclude that the EoR observational windows 
need to be in regions with a very low polarized foreground emission, in order to
minimize `leaked' foregrounds. Faraday rotation measure synthesis, 
polarization surveys obtained by different radio telescopes, and a multiple EoR
observations will help in mitigating the polarization leakages. However, further
simulations and observations are necessary to pin point the best strategy for the
EoR detection.

\section*{acknowledgement}
We acknowledge discussion with the LOFAR-EoR key project members. 
We are also thankful to the anonymous referee for his illustrative and 
constructive comments. As LOFAR members authors are partly funded by 
the European Union, European Regional Development Fund, and by 
`Samenwerkingsverband Noord-Nederland', EZ/KOMPAS.

\bibliographystyle{mn2e}
\bibliography{reflist}

\appendix

\bsp

\label{lastpage}

\end{document}